\documentclass[10pt,twocolumn,letterpaper]{article}

\usepackage{cvpr}              

\pdfoutput=1
\usepackage[dvipsnames]{xcolor}

\usepackage{booktabs, multirow} 
\usepackage{soul}
\usepackage{xcolor}
\usepackage{colortbl} 
\usepackage{changepage,threeparttable} 

\usepackage[frozencache=true,cachedir=minted-cache]{minted}

\definecolor{LightGray}{gray}{0.9}
\definecolor{DraculaBG}{HTML}{282A36}

\definecolor{cvprblue}{rgb}{0.21,0.49,0.74}
\usepackage[pagebackref,breaklinks,colorlinks,allcolors=cvprblue]{hyperref}
\usepackage[accsupp]{axessibility}  

\def\paperID{828} 
\def\confName{CVPR}
\def\confYear{2025}

\title{Design2GarmentCode: \\ Turning Design Concepts to Tangible Garments Through Program Synthesis}

\author{
    \textbf{\small{
        Feng Zhou$^{1,2}$\thanks{Works done during the internship at Style3D Research.} \hskip1.2em
        Ruiyang Liu$^{2,4}$\thanks{Corresponding author.}\hskip1.2em
        Chen Liu$^{2,4}$\hskip1.2em
        Gaofeng He$^{2}$ \hskip1.2em
        Yong-Lu Li$^{3}$\hskip1.2em
        Xiaogang Jin$^{4}$\hskip1.2em
        Huamin Wang$^{2}$
    }}
    \\
    {\small $^{1}$Zhejiang Sci-Tech University} \quad
    {\small $^{2}$Style3D Research} \quad
    {\small $^{3}$Shanghai Jiao Tong University} \quad
    {\small $^{4}$State Key Lab of CAD\&CG, Zhejiang University}
    \\
}

\begin{document}
\maketitle
\begin{abstract}
Sewing patterns, the essential blueprints for fabric cutting and tailoring, act as a crucial bridge between design concepts and producible garments.
However, existing uni-modal sewing pattern generation models struggle to effectively encode complex design concepts with a multi-modal nature and correlate them with vectorized sewing patterns that possess precise geometric structures and intricate sewing relations.
In this work, we propose a novel sewing pattern generation approach \textbf{Design2GarmentCode} based on Large Multimodal Models (LMMs), to generate parametric pattern-making programs from multi-modal design concepts.
LMM offers an intuitive interface for interpreting diverse design inputs, while pattern-making programs could serve as well-structured and semantically meaningful representations of sewing patterns, and act as a robust bridge connecting the cross-domain pattern-making knowledge embedded in LMMs with vectorized sewing patterns.
Experimental results demonstrate that our method can flexibly handle various complex design expressions such as images, textual descriptions, designer sketches, or their combinations, and convert them into size-precise sewing patterns with correct stitches. Compared to previous methods, our approach significantly enhances training efficiency, generation quality, and authoring flexibility. Project page: \href{https://style3d.github.io/design2garmentcode}{https://style3d.github.io/design2garmentcode}.
\end{abstract}    
\begin{figure}
    \centering
    \includegraphics[width=\linewidth]{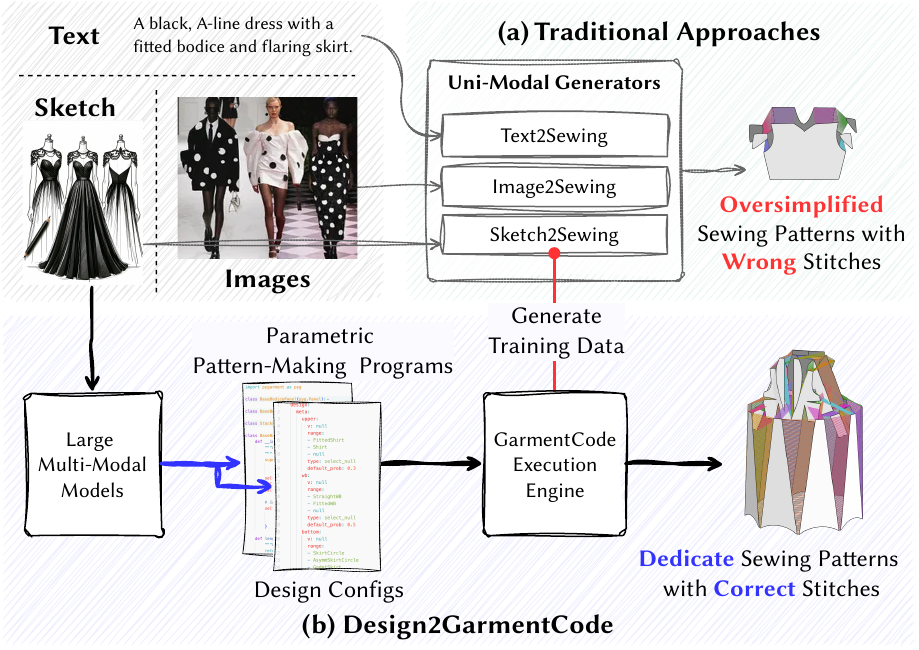}
    \caption{Traditional sewing pattern generation approaches (top) use \textit{uni-modal} models trained on synthetic datasets generated by parametric pattern-making programs (\textcolor[HTML]{FF3333}{red arrow}) to convert text or image prompts into vector-quantized patterns. These methods are resource-intensive and often yield oversimplified patterns with stitching errors. 
    Our approach (bottom) utilizes large pre-trained LMMs to \textit{directly} translate design concepts into parametric programs and configuration files (\textcolor[HTML]{3333FF}{blue arrow}), enabling dedicated, structurally correct pattern generation from multi-modal design inputs within a unified framework.}
    \label{fig:teaser}
\end{figure}

\section{Introduction}
\label{sec:intro}
While generative AI has significantly propelled creativity in fashion design, turning those design ideas into wearable realities remains a formidable challenge. Sewing patterns are the key components to bridge the gap between abstract design ideas and wearable realities. They are foundational blueprints that dictate the precise shapes and dimensions of fabric pieces, essential for assembling garments in both the physical and virtual fashion realms. 

Traditionally, sewing patterns are drafted manually by professional pattern-makers with years of practice, making the process inefficient, error-prone, and unable to meet the growing demands for refinement and personalization in the fashion market. To this end, parametric pattern-making researches~\citep{bao20213d,kang2021development,GarmentCode2023} and industrial solutions~\citep{Valentina,Modaris,Assyst,Optitex} have emerged. These methods formalize the pattern-making process as geometric functions governed by parameters such as body measurements and design features, thereby accelerating the process by enabling pattern makers to generate sewing patterns through parameter adjustments instead of starting from scratch. However, creating these function templates is still complex, requiring not only advanced pattern-making skills but also geometric intuition, mathematical modeling knowledge, and coding abilities to translate pattern-making expertise into CAD programs. These technical barriers significantly restrict the widespread adoption of parametric pattern-making solutions.

\begin{figure}
    \centering
    \includegraphics[width=\linewidth]{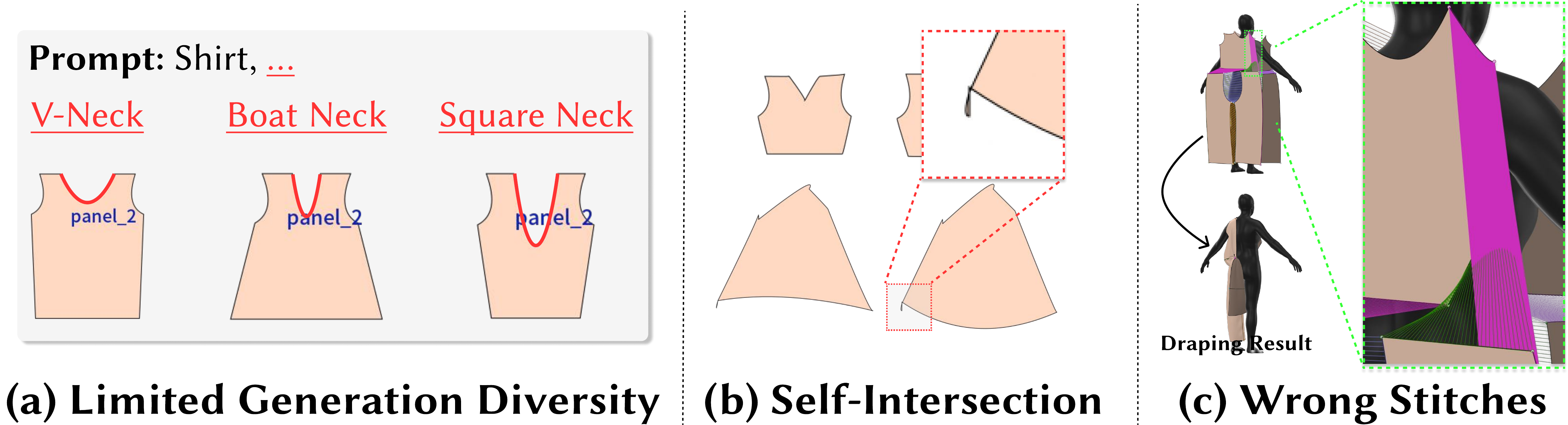}
    \caption{(a) Despite prompts specifying diverse neckline types, DressCode~\citep{he2024dresscode} consistently produces only V-neck designs, indicating limited generation diversity. (b) SewFormer~\citep{liu2023sewformer} often generates sewing patterns with self-intersecting panels, compromising pattern validity. (c) Stitching errors are also prevalent in Sewformer~\citep{liu2023sewformer}, as shown here where a pant side seam is mistakenly stitched to a shirt shoulder seam, resulting in draping failure.}
    \label{fig:prev-problems}
\end{figure}

Recently, several learning-based approaches for sewing pattern generation have been introduced. For instance, NeuralTailor~\citep{korosteleva2022neuraltailor} focuses on extracting sewing patterns from unstructured point clouds, DressCode~\citep{he2024dresscode} targets text-to-sewing pattern generation, and SewFormer~\citep{liu2023sewformer} is designed for image-based sewing pattern generation. 
However, they are generally trained on paired design-sewing pattern data, necessitating large datasets to effectively capture the multi-modal nature of design concepts. Furthermore, sewing patterns require centimeter-level precision to ensure proper garment fit, which presents a significant challenge for neural networks that only provide statistical approximations of the true function based on their training data ~\citep{henighan2020scaling,duan2022comprehensive,ritchie2023neurosymbolic}. As a result, these methods frequently generate oversimplified patterns with flawed geometry or stitches, potentially leading to draping failures (Figure~\ref{fig:prev-problems}).

In this paper, we present \textbf{Design2GarmentCode}, an innovative approach that leverages the generalization capabilities of vision-language foundation models to achieve multi-modal sewing pattern generation with \textit{minimal computational and data requirements}. Unlike previous methods that directly synthesize vector-quantized patterns, Design2GarmentCode employs LMMs to learn the syntax of parametric pattern-making programs, translating design concepts into \textit{parameters and programs} that can be executed to produce precise and structurally accurate sewing patterns.
Design2GarmentCode combines a pre-trained Large Multimodal Model (LMM) as a design interpreter with a finetuned Large Language Model (LLM) as a program synthesizer. Specifically, the LLM is finetuned on code snippets from GarmentCode~\citep{GarmentCode2023}, a domain-specific language for constructing parametric sewing patterns.
At runtime, the design interpreter extracts both topological and geometrical information from the design input by responding to a series of questions from the program synthesizer, which then generates garment programs and design configurations following GarmentCode syntax. Our method offers the following major contributions:

\begin{itemize}
    \item We introduce a novel \textbf{modality-agnostic} framework with an intuitive, intelligent interface capable of processing user design intentions across multiple modalities simultaneously by integrating pre-trained LMMs.
    
    \item We present the \textbf{first} sewing pattern generation approach grounded in \textbf{program synthesis}, delivering \textit{fully interpretable}, \textit{geometrically precise}, and \textit{structurally accurate} patterns through a more compact, semantically clear, and LLM-friendly representation.
    
    \item Our framework \textbf{benefits real-world production} by enabling flexible pattern authoring through natural language or physical feedback, allowing precise customization and efficient \textit{creation of novel garment components}, represented as parametric pattern-making programs.
    
    \item Our approach requires only minimal fine-tuning of a pre-trained LLM and the training of a lightweight, text-conditioned transformer decoder, making it \textbf{more efficient} than existing vector-quantized sewing pattern generation models trained from scratch while offering \textit{superior generation quality and authoring flexibility}.
\end{itemize}
\section{Related Work}

\subsection{Garment Modeling with Sewing Patterns} 
Garment modeling and generation can be broadly classified into two categories: direct 3D garment generation (meshes) and sewing pattern generation, which are later draped onto human bodies via cloth simulation~\citep{liu2024automatic,wang2018parallel,wang2018rule} or learning-based techniques~\citep{li2024diffavatar,li2024isp}. 
Direct 3D garment generation often relies on differentiable garment representations like unsigned distance fields~\citep{zhou2024udiff,yu2023surf}, shells~\citep{liu2023gshell}, or Gaussian splatting~\citep{rong2024gaussian}. However, it presents challenges in terms of both geometric accuracy and editability. On one hand, capturing fine garment details like folds and wrinkles necessitates extremely high-resolution 3D representations. On the other hand, editing these generated garments requires a well-defined UV space, and flattening the 3D mesh into developable meshes demands careful consideration from both geometric and statistical perspectives~\citep{pietroni2022computational,bang2021estimating}.

Sewing pattern generation, by contrast, has been approached through both learning-based and procedural modeling methods. Learning approaches utilize vector-quantized representations of sewing patterns, mapping from unstructured 3D point clouds~\citep{bang2021estimating,korosteleva2022neuraltailor}, images~\citep{jeong2015garment,chen2022structure,liu2023sewformer,yang2018physics}, or textual descriptions~\citep{he2024dresscode} to structured patterns. However, they are highly dependent on the quality and diversity of the training data and often struggle to generalize designs beyond the training domain. Furthermore, generating high-quality 3D garment data requires substantial domain knowledge and the involvement of skilled professionals.

Procedural modeling is an alternative that relies on predefined rules and parameters to generate garment patterns. For instance, GarmentCode~\citep{GarmentCode2023}, a DSL for parametric pattern making, enables precise control over garment design and customization. The GarmentCodeData~\citep{GarmentCodeData:2024}, built on GarmentCode, further illustrates the potential of procedural methods to generate a diverse range of made-to-measure garments, with adaptability to different body shapes. While procedural modeling provides greater control and precision, it typically requires specialized expertise and is less flexible when dealing with novel or unconventional designs.

\subsection{LLMs for Program Synthesis}
Recent advancements in program synthesis and code generation using large language models (LLMs) have laid essential groundwork for systems like Design2GarmentCode, which generate structured garment code from multi-modal design inputs. Earlier researches like Codex~\citep{chen2021evaluating} and AlphaCode~\citep{li2022competition} demonstrated the effectiveness of LLMs in generating complex, task-specific code with high syntax accuracy, showcasing potential in scenarios requiring precise parametric coding. These models~\citep{nijkamp2022codegen,xu2022polycoder} highlight how LLMs, when sufficiently trained, can transform natural language inputs into executable code, a capability directly relevant to generating garment codes that follow complex pattern-making syntax. ~\citep{wang2021codet5,fried2022incoder,bassamzadeh2024comparative,gal2024comfygen,xue2024genagent} explore multi-modal models that integrate visual aids, such as flowcharts and UML diagrams, into LLM training to enhance models’ comprehension of complex structures and flow. These models particularly emphasize the need for semantic understanding and adaptability, which are critical in working with domain-specific languages (DSLs) like GarmentCode.

\subsection{Neurosymbolic Models}

Procedural/symbolic models and learned/neural models have complementary strengths and weaknesses. Neurosymbolic models~\citep{ritchie2023neurosymbolic} tend to combine the strengths of both paradigms and propose to generate visual data using symbolic programs augmented with AI/ML techniques. The neurosymbolic pipeline typically includes task specification, program synthesis using a DSL, program execution, and optional neural post-processing for refining results. It has been successfully applied across several areas of computer graphics. 
In 2D shape modeling, they are used in layout generation~\citep{wang2019planit,para2021generative}, engineering sketch creation~\citep{edward1963sketchpad,para2021sketchgen,seff2021vitruvion}, and vector graphics synthesis by constructing programs that represent geometric shapes and their spatial relations~\citep{carlier2020deepsvg,ellis2021dreamcoder}. 
In 3D shape modeling, they facilitate inferring shape programs from existing 3D models~\citep{tian2019learning,li2020sketch2cad,jones2022plad,jones2023shapecoder} or generating entirely new 3D shapes by training generative models on shape programs~\citep{wu2021deepcad,xu2022skexgen,xu2024brepgen} or generate generate node graphs that define complex textures and materials~\citep{shi2020match,guerrero2022matformer,hu2022inverse} following the procedural modeling paradigm. Additionally, neuro-symbolic methods have been employed in human motion prediction~\citep{feng2024chatpose,li2024isolated}, reasoning~\citep{wu2024symbol,li2024labyrinth,zhang2025take} and generation~\citep{chen2023executing,liu2024primitive,xu2024humanvla}, which leveraging visual-language foundation models to extract symbolic representations from visual data, facilitating complex activity reasoning by combining visual cues with symbolic logic~\citep{wu2024symbol}.
Our method leverages a neurosymbolic approach by instruction-tuning LLMs to generate GarmentCode design configurations and component programs from diverse design concepts, ensuring geometric and structural accuracy.

\begin{figure*}[t]
    \centering
    \includegraphics[width=\linewidth]{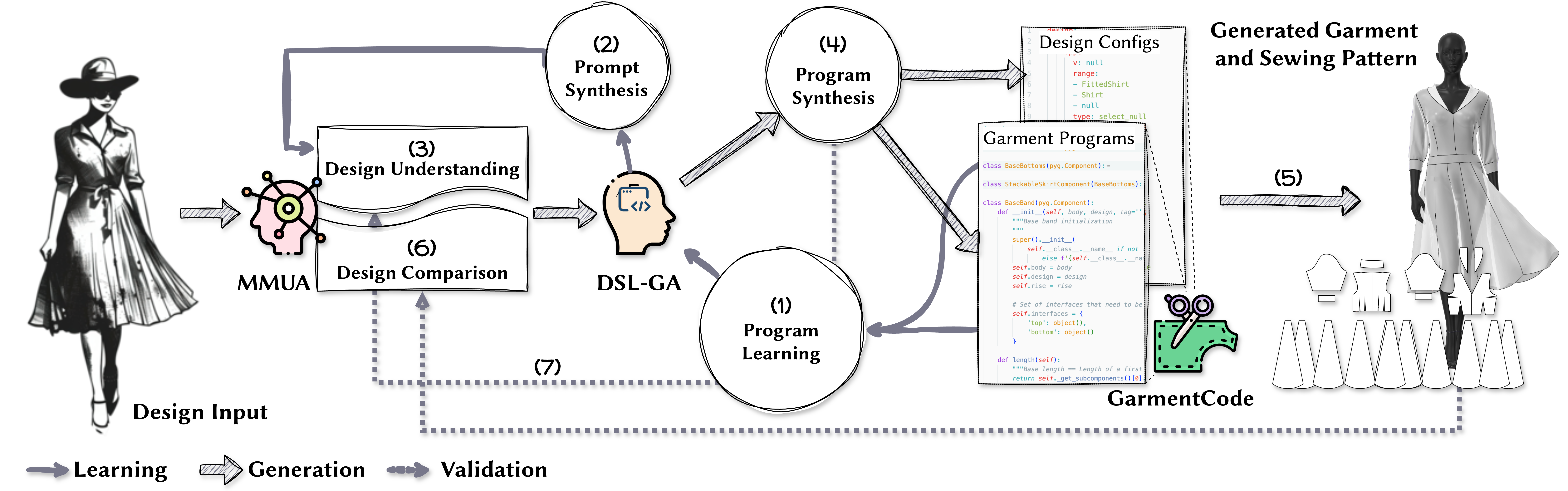}
    \caption{Overview of Dress2GarmentCode. (1) \textbf{Program Learning}: we finetune the \textit{DSL Generation Agent (DSL-GA)} using GarmentCode example programs, teaching it the GarmentCode grammar and the semantics of each design parameter. (2) \textbf{Prompt Synthesis}: the DSL-GA generates prompts for the \textit{Multi-Modal Understanding Agent (MMUA)} to interpret and extract relevant design features from the input (3). (4) \textbf{Program Synthesis}: based on the MMUA's responses, the DSL-GA synthesizes GarmentCode-compliant design configurations and garment programs, which are then executed by the GarmentCode engine to produce sewing patterns and simulated garments (5). To enhance robustness, we incorporate two validation loops: during program synthesis, we employ rule-based validations (7) to ensure the MMUA's outputs are sufficient for generating complete and valid garment programs and design parameters; after the initial generation, the MMUA compares the generated design with the input and suggests modifications to minimize discrepancies.}
    \label{fig:pipeline}
\end{figure*}

\section{Method}
Our goal is to develop a generative model that transforms multi-modal design concepts into precise sewing patterns. This requires understanding diverse inputs and producing patterns with high geometric precision and intricate structures. 
These requirements present a challenge for conventional models, which require
extensive training data and struggle with output precision due to their probabilistic nature. 
We propose \textbf{Design2GarmentCode}, a system leveraging LMMs to \textbf{generate parametric pattern-making programs}, or specifically GarmentCode~\citep{GarmentCode2023}.
Design2GarmentCode reduces the need for large datasets utilizing the pre-embedded pattern-making knowledge in LMMs while ensuring output precision with parametric program synthesis. In the following, we first provide an overview of parametric pattern-making programs and GarmentCode syntax, and then describe the detailed design of Design2GarmentCode.

\subsection{Parametric Sewing Patterns}
\paragraph{Parametric sewing patterns} are formally represented as symbolic programs that generate sewing patterns (i.e., 2D CAD sketches) based on body measurements and design configurations. These symbolic programs enhance the efficiency of the pattern-making process by allowing users to draft or modify sewing patterns through semantically meaningful parameters. Mathematically, we can represent a sewing pattern $\mathcal{S}$  as:
\begin{equation}
    \mathcal{S} = \langle \mathcal{F}, \mathcal{D}, B \rangle = \cup_{f_i\in\mathcal{F},d_i\in\mathcal{D}} f_i(d_i, B),
\end{equation}
where $\mathcal{F}$ is the set of symbolic programs, $\mathcal{D}$ represents design configurations, and $B$ represents body measurements. Each symbolic function $f_i\in\mathcal{F}$ is essentially a series of rule-based 2D draw-calls controlled by its unique set of design configurations $d_i\in\mathcal{D}$ and the body measurements $B$.

\paragraph{GarmentCode} is a domain-specific language (DSL) designed to generate parametric sewing patterns by encapsulating those symbolic programs in a hierarchical, component-oriented manner. In GarmentCode, each symbolic program $f_i$ uses parametric curves to define a garment component, such as sleeves, bodices, or collars. The smallest component is a single panel, and multiple components can be combined through interface functions to create a larger component. 
In GarmentCode, a complete sewing pattern is specified by topological parameters $\mathcal{D}_T$ (which define the presence and quantity of garment components) and geometrical parameters $\mathcal{D}_G$, which determine the dimensions of each component when combined with body measurements $B$. 
As this work primarily focuses on design variations, we use a standard body model throughout all experiments to ensure consistency.

\subsection{The Design2GarmentCode System}
\label{sec:lmm_agents}
As illustrated in Figure~\ref{fig:pipeline}, Design2GarmentCode has three components: \textbf{DSL Generation Agent (DSL-GA)}, a finetuned LLM responsible for (1) program learning, (2) prompt synthesis, and (4) program synthesis; \textbf{Multi-modal Understanding Agent (MMUA)}, a pre-trained LLM that manages design understanding (3) and design comparison (6); and (5) \textbf{GarmentCode}, which executes the synthesized programs to generate sewing patterns and 3D garments.

The system workflow begins with \textbf{Program Learning} (Sec.~\ref{sec:prog_learning}), where DSL-GA is finetuned to understand the syntax and semantic meanings of GarmentCode parameters. In \textbf{Prompt Synthesis}, DSL-GA creates prompts for MMUA to identify essential design features. These features are then provided to DSL-GA for \textbf{Program Synthesis}, where garment programs and design configurations are generated through rule-based parameter validation (Figure~\ref{fig:pipeline} (7)) and a learned projector (Sec.~\ref{sec:prog_synthesis}). The GarmentCode Execution Engine then produces sewing patterns and draped garment models. Finally, a \textbf{Validation} stage compares the generated garment with the original design, allowing MMUA to provide specific correction instructions to DSL-GA for iterative refinement, such as \textit{``make the sleeve longer''}.

\subsubsection{Program Learning}
\label{sec:prog_learning}
During experiments, we found that pre-trained LLMs have some foundational knowledge of pattern drafting. For example, when prompted with “How to draft a basic upper body bodice?”, LLMs can produce drafting instructions that align with conventional practices. We use a pre-trained LLM to initialize DSL-GA, however, due to GarmentCode's customized object notations and function logistics, directly prompting DSL-GA to generate GarmentCode programs poses significant challenges~\citep{bassamzadeh2024comparative}.

To address these challenges, we propose to align DSL-GA's embedded pattern-making knowledge with the specific syntax and semantics of GarmentCode via LoRA~\citep{hu2021lora} based on fine-tuning. We start by providing the DSL-GA (denoted as $\Gamma$) with existing GarmentCode programs $\mathcal{F}$, and instructing it to comment on the functions with detailed pattern-drafting instructions. After manually validating the comments, we get a dataset $D$ paring natural language instructions with GarmentCode implementations:
\begin{equation}
  D = \bigg\{\big( \Gamma_{cmt}(f_i), f_i \big) \,\ | \,\, f_i\in \mathcal{F} \bigg\},  
\end{equation}
where $\Gamma_{cmt}$ is the instructed DSL-GA for code commenting, $f_i$ is . Similar to \citep{bassamzadeh2024comparative}, we finetune DSL-GA ($\Gamma$) on the dataset $D$ with LoRA~\citep{hu2021lora}, aiming for $\Gamma_{ft}(\Gamma_{cmt}(f_i))\rightarrow f_i$, where $\Gamma_{ft}$ is the finetuned DSL-GA. 

After fine-tuning, the fine-tuned DSL-GA $\Gamma_{ft}$ gains an understanding of the code structure and parameter semantics in GarmentCode (Figure~\ref{fig:app_new_code}). Therefore, we provide the design configuration $\mathcal{D}$ to $\Gamma_{ft}$, prompting it to analyze the semantic meaning of each parameter and generate structured queries. These queries are designed to guide the MMUA in extracting relevant design features from the multi-modal input, enabling $\Gamma_{ft}$ to generate a comprehensive set of design parameters. The generated prompt $P$ typically starts with analysis instructions, followed by multiple-choice or numerical estimation questions regarding each design parameter $d_i$. Formally, we have
\begin{equation}
 P=\Gamma_{ft}(\mathcal{D})=\cup_{d_i\in\mathcal{D}}\Gamma_{ft}(d_i)=\cup q_i,   
\end{equation}
where $q_i=\Gamma_{ft}(d_i)$ represents the generated question regarding the $i$-th design parameter $d_i$.

\subsubsection{Program Synthesis}
\label{sec:prog_synthesis}
Initial results showed that MMUA performed significantly better on multiple-choice questions compared to numerical estimation questions. To improve accuracy, we replaced all numerical estimation questions in the initial prompt $P$ with equivalent \textbf{multiple-choice questions} with descriptive options such as \textit{``full length''}, \textit{``half length''}, or \textit{``three-quarter length''}. We append a lightweight \textbf{projector} $\Psi$ after the finetuned DSL-GA $\Gamma_{ft}$ to transform these descriptive answers $\tau_i$ regarding the design input $x$ into precise geometrical parameters $d_i\in\mathcal{D}$ adhering to GarmentCode~\citep{GarmentCode2023}:
\begin{equation}
    \Psi: \Gamma_{ft}(\cup\tau_i)\rightarrow \mathcal{D}, \, \text{ where }\tau_i=\textbf{MMUA}(q_i,x).
    \label{eq:ft_text_enc}
\end{equation}

Inspired by DressCode~\citep{he2024dresscode}, we implement the projector $\Psi$ as text-conditioned decoder-only transformer, where we design a \textbf{type-based quantization function} $\mathbf{Q}$ to convert the parameter list $\mathcal{D}$ into a token sequence $\mathcal{T}=\{t_1,...,t_N\}$, where $N=|\mathcal{D}|$ denote the total number of design parameters. The quantization function $\mathbf{Q}$ operates as follows:
\begin{equation}
    t_i=\mathbf{Q}(d_i) =
    \begin{cases}
        0/1, & \text{if } d_i \text{ is a boolean variable}, \\
        d_i, & \text{if } d_i \text{ is an integer}, \\
        \lambda \cdot \textbf{Norm}(d_i), & \text{if } d_i \text{ is a floating number}, \\
        \textbf{Index}(d_i, L), & \text{if } d_i \text{ is a selective variable}. \\
    \end{cases}
    \label{eq:quantize_d_param}
\end{equation}
where $\lambda$ is a scaling factor indicating numerical precision. We use $\lambda=100$ to maintain centimeter-level precision.

As in Eq.~\ref{eq:ft_text_enc}, we use the finetuned \textbf{DSL-GA} $\Gamma_{ft}$ to encode the answers $\tau_i$ from \textbf{MMUA} and construct the condition input for $\Psi$ (we use an MLP to match the embedding dimension between $3,072$ in $\Gamma_{ft}$ and $128$ in $\Psi$). Notably, our token sequence length is fixed to the number of design parameters $|\mathcal{D}|=122$, regardless of the complexity of the pattern. This fixed-length representation is at least $10\times$ compact than DressCode~\citep{he2024dresscode}, whose sequence length is $1,500$ and scales with pattern complexity (see Sec.~\ref{supp:implementation} for implementation details). 
\begin{table}[t]
\centering
\resizebox{\linewidth}{!}{
\begin{tabular}{cl|cc|cc}\toprule
\multicolumn{2}{c|}{\multirow{2}{*}{\textbf{Method}}} &\multicolumn{2}{c|}{\textbf{\quad\quad Text Guided Generation \quad\quad}} &\multicolumn{2}{c}{\textbf{\quad\quad Image Guided Generation \quad\quad}} \\
& &DressCode~\citep{he2024dresscode} &Ours &Sewformer~\citep{liu2023sewformer} &Ours \\\midrule
\multirow{6}{*}{\textbf{Quality}} &SSR &84\% &\cellcolor[HTML]{f1f0fc}100\% &65.33\% &\cellcolor[HTML]{f1f0fc}94\% \\
&Agreement & 7.17\% & \cellcolor[HTML]{f1f0fc}79.83\% & 3.33\% & \cellcolor[HTML]{f1f0fc}88.67\% \\
&Aesthetic & 9.50\% & \cellcolor[HTML]{f1f0fc}68.17\% & 5.33\% & \cellcolor[HTML]{f1f0fc}77\% \\
&CLIPScore & 0.2386 & \cellcolor[HTML]{f1f0fc} 0.2438 & / & / \\
&F-Score & 0.616 & / & 0.708 & \cellcolor[HTML]{f1f0fc}0.829 \\
&CD & 15.77 & / & 9.7 & \cellcolor[HTML]{f1f0fc}2.091 \\
\midrule
\multirow{3}{*}{\textbf{Diversity}} 
&\# Panels &$5.11_{\pm 1.66}$ &\cellcolor[HTML]{f1f0fc}$6.92_{\pm 3.10}$ &\cellcolor[HTML]{ececec}$10.11_{\pm \textbf{4.42}}$ &\cellcolor[HTML]{ececec}$\textbf{11.02}_{\pm 4.18}$ \\
&\# Edges &$5.48_{\pm 1.60}$ &\cellcolor[HTML]{f1f0fc}$6.84_{\pm 3.38}$ &$5.79_{\pm 1.71}$ &\cellcolor[HTML]{f1f0fc}$6.24_{\pm 2.90}$ \\
&\# Stitches &$10.06_{\pm 3.24}$ &\cellcolor[HTML]{f1f0fc}$18.66_{\pm 8.64}$ &$15.81_{\pm 5.91}$ &\cellcolor[HTML]{f1f0fc}$27.9_{\pm 9.83}$ \\
\bottomrule
\end{tabular}}
\caption{Quantitative comparison of our method against state-of-the-art (SOTA) sewing pattern generation techniques in terms of quality and diversity. 
\textit{SSR} (Simulation Success Rate) indicates the feasibility of simulated garment assembly, while \textit{Agreement} measures alignment with design prompts, and \textit{Aesthetic} evaluates the visual preference of the generated patterns. \textit{CLIPScore} assesses text-image consistency, whereas \textit{Chamfer Distance (CD)} and \textit{F-Score} quantify geometric accuracy. \textit{\#Panels}, \textit{\#Stitches}, and \textit{\#Edges} denote the mean and standard deviation (subscript) of the number of panels, stitches per pattern, and edges per panel.}
\label{tab:quant_eval}
\end{table}

\begin{figure*}[t]
    \centering
    \includegraphics[width=\linewidth]{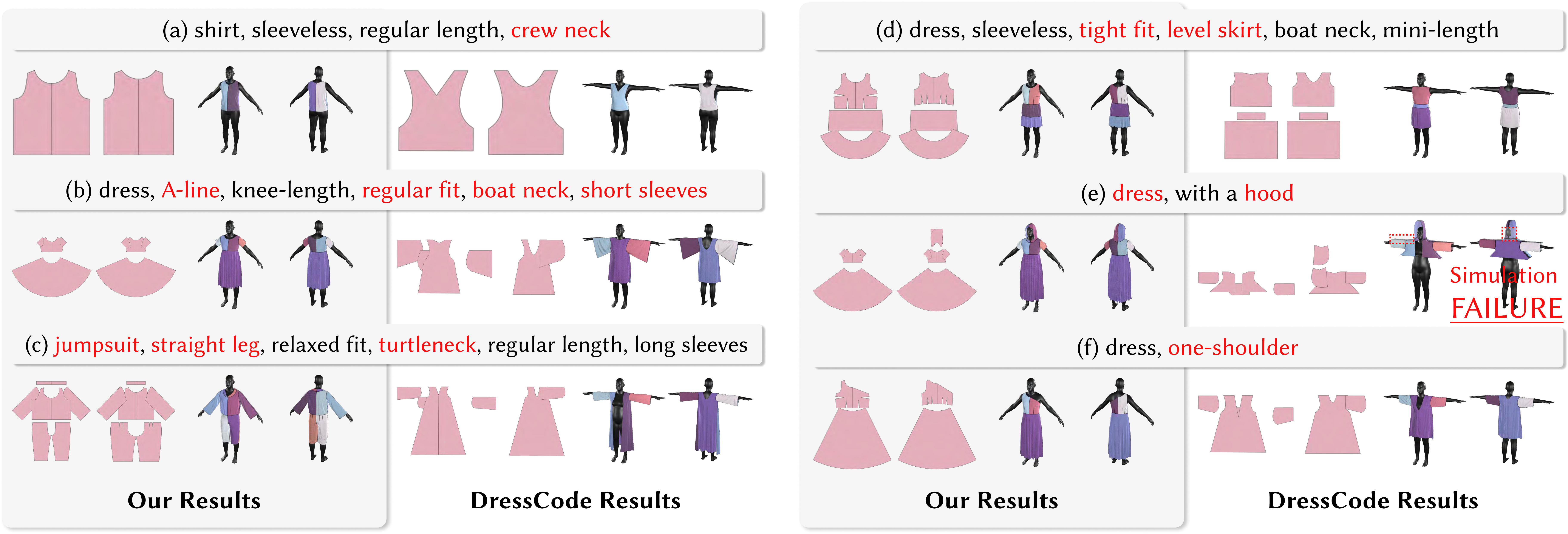}
    \caption{Quality Comparison on Text-Guided Sewing Pattern Generation. For each design, we present the generated pattern using our method (left) alongside DressCode~\citep{he2024dresscode} (right), including front and back renderings of the draped garment. We highlight design elements accurately captured by our method but missed by DressCode~\citep{he2024dresscode} use red color in the input prompt.}
    \label{fig:result_text}
\end{figure*}

\section{Experiments}
\subsection{Quantitative Evaluation}
\label{sec:quant_eval}
We evaluate our proposed method against state-of-the-art sewing pattern generation approaches (DressCode~\citep{he2024dresscode} for text-guided and Sewformer~\citep{liu2023sewformer} for image-guided generation) on \textbf{Generation Quality} and \textbf{Generation Diversity}.

\textbf{Generation Quality} is evaluated through three metrics: \textit{Simulation Success Rate (SSR)}, \textit{ Agreement Score}, and \textit{Aesthetic Score}. The \textit{Simulation Success Rate (SSR)} is calculated as the ratio of successfully simulated garments to the total number of generated sewing patterns, measuring the structural feasibility of the patterns. We prepared a dataset comprising 150 text prompts and 150 test images, covering a wide variety including \textit{tops (78), pants (76), skirts (38), dresses (80), and suits (28)}. For each sample, we generated sewing patterns using both our method and baseline methods, and simulated the patterns using GarmentCode’s simulation engine~\citep{GarmentCode2023,GarmentCodeData:2024} to compute the success rate. The \textit{Agreement and Aesthetic Scores} were derived from a user study involving 30 professional pattern-makers. Each participant is asked to review 50 text and 50 image samples generated by our and the baseline models, and assess their preference based on sewing patterns and simulated garments according to:
\begin{itemize}
    \item \textit{Agreement}: the degree to which the generated pattern matched the design prompt.
    \item \textit{Aesthetic Quality}: the visual appeal and structural coherence of the generated pattern.
\end{itemize}
For each criterion, participants could express a preference for either our method or the baseline or indicate that both methods were \textit{``comparable''}. We calculate the Agreement and Aesthetic Scores as the percentage of times each option was chosen over the total number of tested samples.

Table~\ref{tab:quant_eval} presents the results, showing that our method surpasses existing approaches in both SSR and user-evaluated Agreement and Aesthetic scores. For text-guided generation, our model achieves a perfect 100\% SSR, notably higher than DressCode’s 84\%. Additionally, our Agreement score of 79.83\% and Aesthetic score of 68.17\% far exceed DressCode’s respective scores of 7.17\% and 9.5\%. In image-guided generation, our method attains a 94\% SSR, with an Agreement score of 88.67\% and an Aesthetic score of 77\%, significantly outperforming Sewformer. These enhancements highlight our model’s ability to generate sewing patterns that are both structurally precise and visually aligned with the design prompt.

\paragraph{Generation Diversity} is evaluated by analyzing the average number of panels (\# Panels), edges (\# Edges), and stitches (\# Stitches) in the generated patterns. For text-guided generation, our method yields more intricate designs, with an average of 6.92 panels, 6.84 edges, and 18.66 stitches per pattern, compared to DressCode's simpler outputs of 5.11 panels, 5.48 edges, and 10.06 stitches. In image-guided generation, our approach also demonstrates superior diversity, producing an average of 11.02 panels, 6.24 edges, and 27.9 stitches per pattern, compared to Sewformer's averages of 10.11 panels, 5.79 edges, and 15.81 stitches. These results emphasize our model’s ability to capture and replicate subtle design variations, highlighting its robustness and adaptability across different design inputs.

\begin{figure*}
    \centering
    \includegraphics[width=\linewidth]{figs/result_img.pdf}
    \caption{Quality Comparison on Image-Guided Sewing Pattern Generation. We compare our method with Sewformer~\citep{liu2023sewformer} on Internet-collected fashion photographs (left), and AI-generated design images without human models (right). The results indicate that our method successfully captures design details from diverse styles, producing sewing patterns that accurately reflect neckline (a, d), cuffs (a, e, g), darts (c, d), and asymmetry (f). In contrast, Sewformer’s results exhibit several issues, including incorrect necklines (a, d), missing components (b, g), misplaced or imaginary stitches (d, e), and extraneous pattern pieces (h). Additionally, since Sewformer’s pattern generation does not account for body shape, garments like skirts and pants frequently appear oversized around the waist, causing them to sag when draped.}
    \label{fig:result_image}
\end{figure*}

\begin{figure*}
    \centering
    \includegraphics[width=\linewidth]{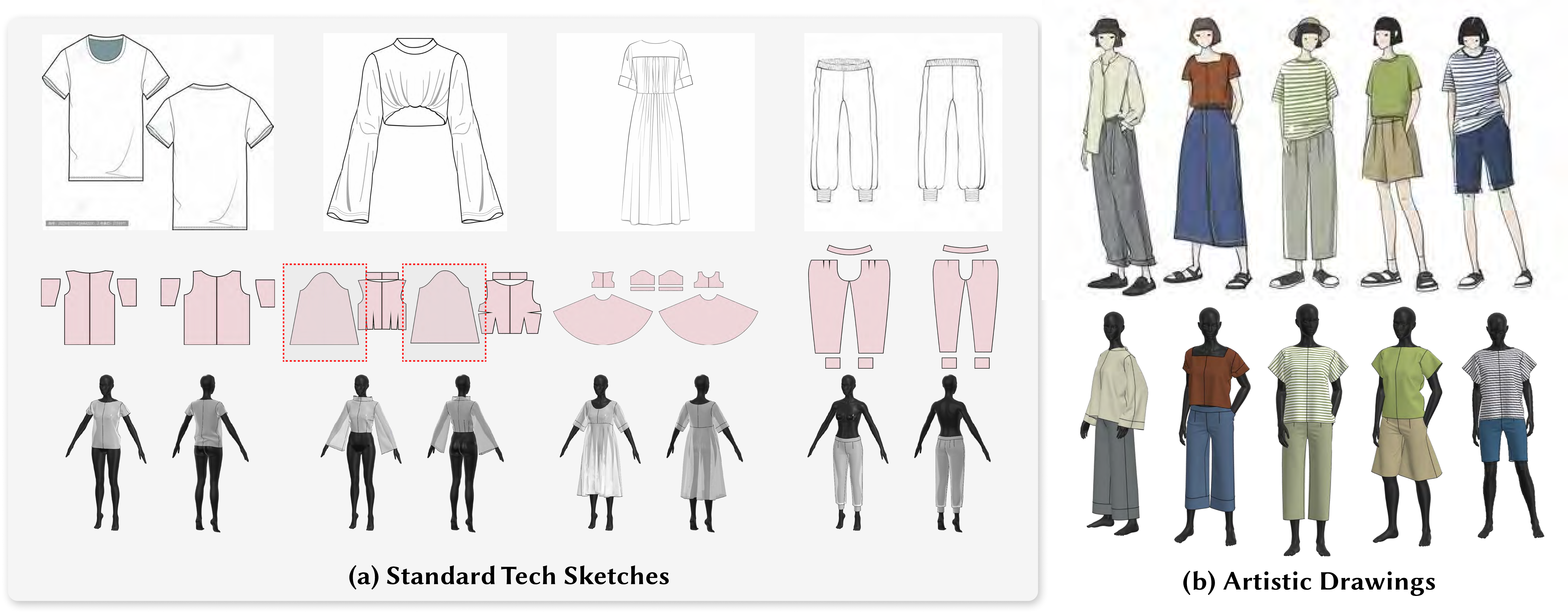}
    \caption{Examples of sketch-based sewing pattern generation. Our method was able to generate high-quality sewing patterns from design sketches under various styles and could integrate seamlessly with industrial fashion design software for (a) pattern editing, i.e. sleeve panels in red boxes are merged from separate front/back sleeve panels; and (b) avatar posture and fabric material editing.}
    \label{fig:result_sketch}
\end{figure*}

\begin{figure*}
    \centering
    \includegraphics[width=0.95\linewidth]{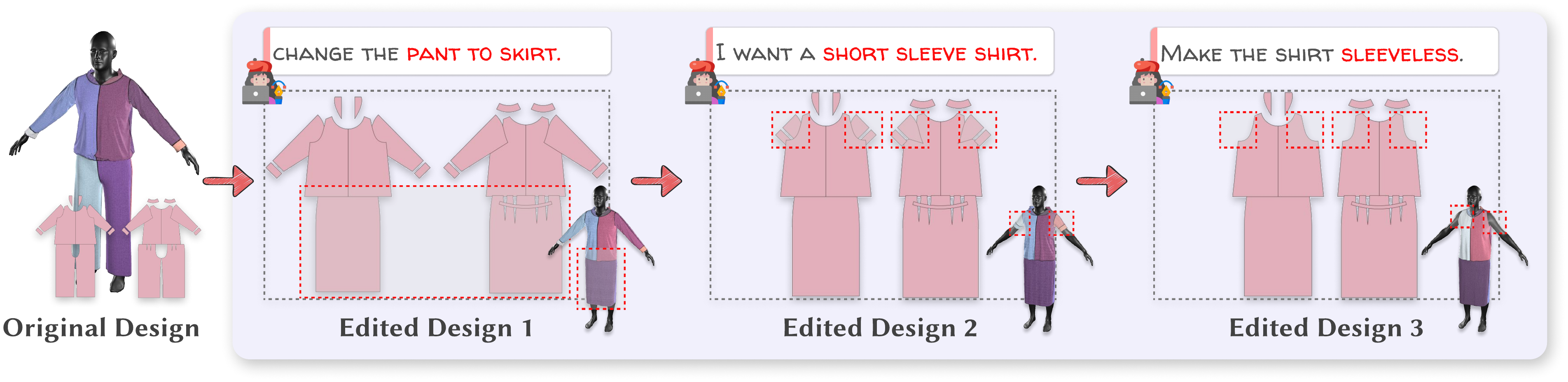}
    \caption{Sewing Pattern Authoring with instructions. Starting from an original design, the system follows user instructions to adjust specific pattern elements. In Edited Design 1, the pants are modified to a skirt based on the command ``CHANGE THE PANT TO SKIRT". In Edited Design 2, the sleeves are shortened as requested. Finally, in Edited Design 3, the shirt is made sleeveless in response to the instructions. Note that, each modification accurately applies only to the specified parts, leaving the rest of the design unchanged.}
    \label{fig:app_editing}
\end{figure*}

\subsection{Multi-modal Generation Results}
Our proposed method demonstrates superior performance across various sewing pattern generation tasks, including text-guided, image-guided, and sketch-based generation.

In text-guided sewing pattern generation (Figure~\ref{fig:result_text}), our method accurately captures design details specified in prompts, such as neckline types (\eg, crew neck (a), boat neck (b), turtleneck (c)) and complex structural features like asymmetry (f) and layered skirts (d). In comparison, the baseline model DressCode struggles with limited pattern diversity, often defaulting to simpler shapes like V-neck designs. Additionally, for design descriptions out of its training domain, DressCode frequently generates patterns with incorrect stitching, leading to poor draping results (Figure~\ref{fig:result_text} (e)). Our method could provide structurally sound and visually accurate patterns under a large design variety, showcasing its capability to handle diverse design requests with high fidelity.

For image-guided sewing pattern generation (Figure~\ref{fig:result_image}), our model effectively translates detailed visual cues from input images into corresponding sewing patterns. Compared with Sewformer, which often fails to model-specific design elements like cuffs, hoods, and asymmetric features, our approach accurately reproduces these details. Sewformer's results frequently exhibit structural flaws, such as missing or misaligned pattern pieces and extraneous components, resulting in unrealistic garment draping. In contrast, our method maintains structural integrity and captures complex design features, producing patterns that closely align with the source images.

In sketch-based sewing pattern generation (Figure~\ref{fig:result_sketch}), our system seamlessly converts both technical sketches (left) and artistic drawings\footnote{The drawing is borrowed from the artwork of \href{https://www.zcool.com.cn/work/ZMzAzNDU0MzY=.html}{TWELVEYIN}.} (right) into high-quality sewing patterns. We also demonstrated that the generated sewing patterns could seamlessly integrate into industrial fashion design software\footnote{We use \href{https://studio.style3d.com/}{Style3D Studio}~\citep{Style3D} for pattern and appearance authoring.}. For example, highlighted sleeve panels in Figure~\ref{fig:result_sketch} (a) are merged from separate front and back sleeve pieces, while Figure~\ref{fig:result_sketch} (b) demonstrates avatar posture and fabric material editing.

\section{Application}
In this section, we explore practical applications enabled by our system that extend beyond basic pattern generation, providing designers with versatile tools for design refinement, integration with physical simulation, and the creation of new garment components.

\textbf{Instruction-Based Editing}. Our system allows designers to adjust generated sewing patterns through simple, instruction-based edits, utilizing the same refinement process as in our collaborative framework. As illustrated in Figure~\ref{fig:app_editing}, starting from an original design, the system responds to natural language commands from the user to adjust the sewing pattern. At each step, the modified areas are highlighted in red boxes. From the figure, it is evident that our system can accurately update only the specified parts of the pattern according to the user's instructions while leaving all other parts of the design unchanged.

\begin{figure}
    \centering
    \includegraphics[width=0.95\linewidth]{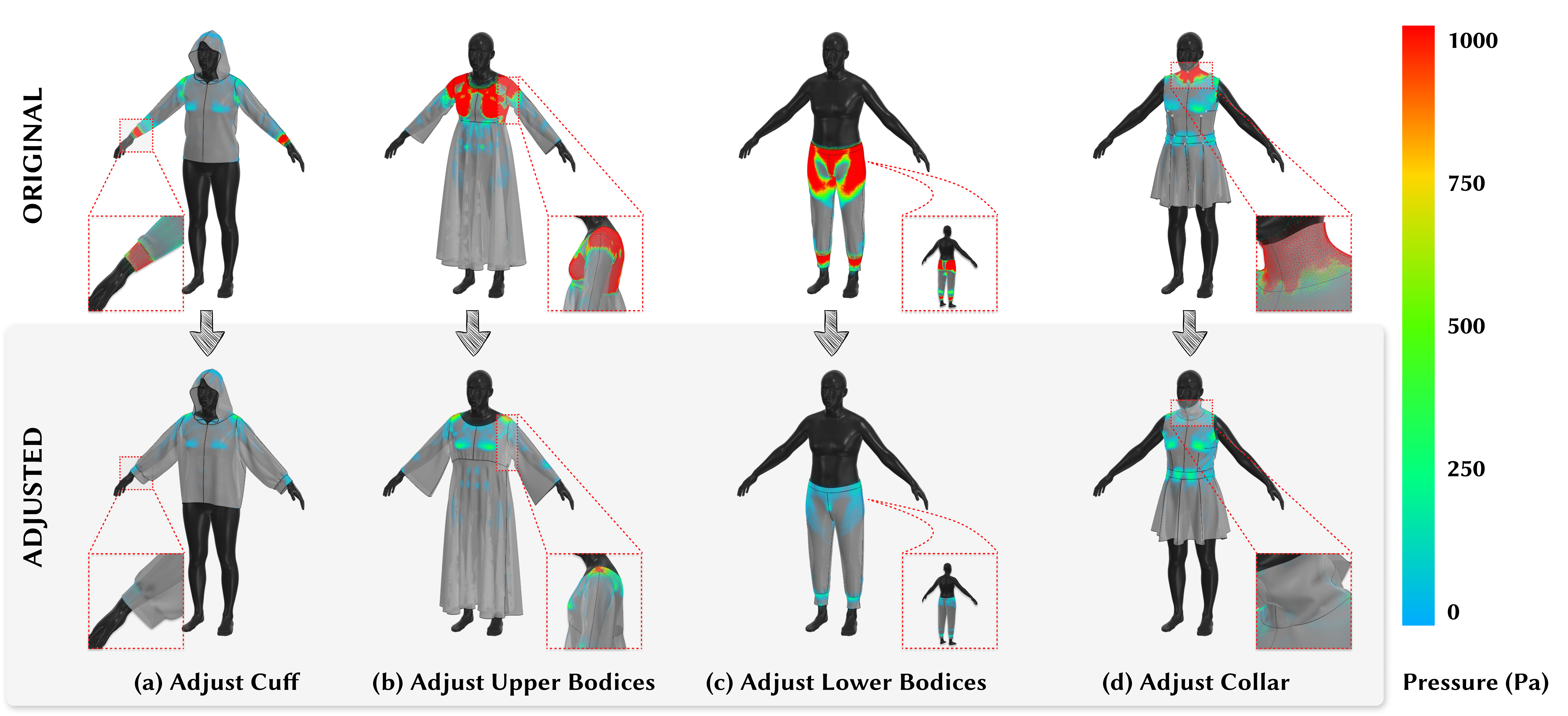}
    \caption{Sewing pattern adjustment based on body pressure measurement. Red regions indicate areas of tight fabric with high body pressure, while blue regions represent looser areas.}
    \label{fig:app_phys}
\end{figure}

\begin{figure}
    \centering
    \includegraphics[width=0.95\linewidth]{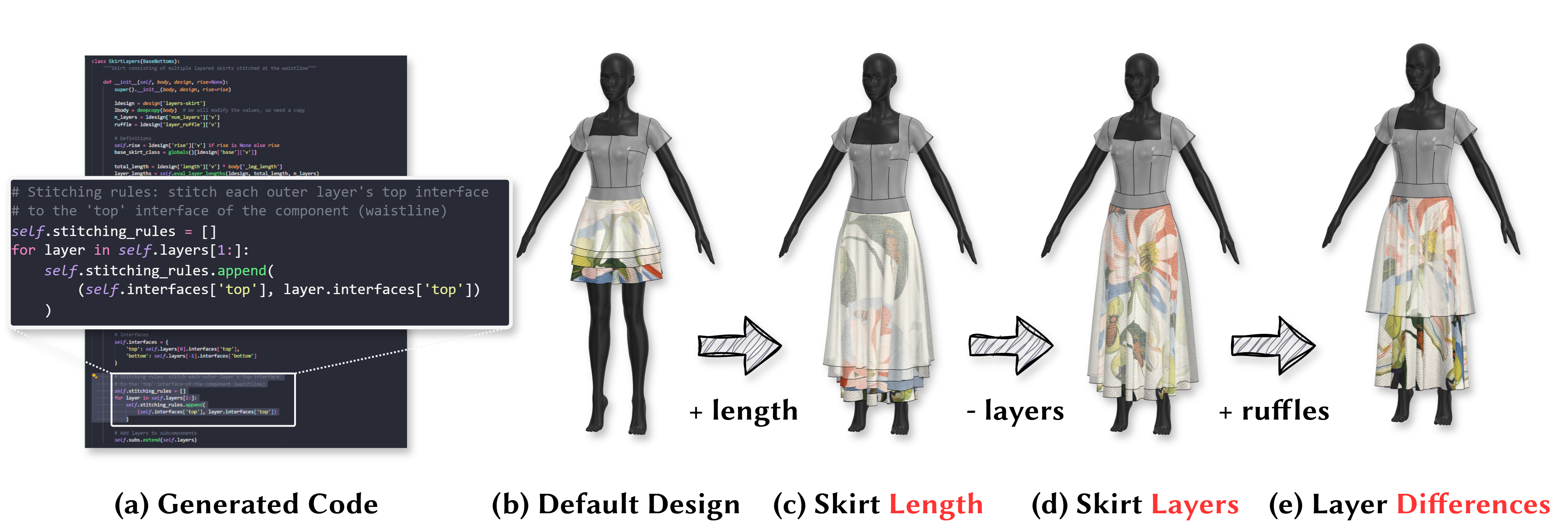}
    \caption{The code for a layered-skirt component generated by our DSL-GA and 3D garment under various design parameters.}
    \label{fig:app_new_code}
\end{figure}

\textbf{Physics-Based Editing}. Our system's generated sewing patterns integrate seamlessly with professional cloth simulation software, allowing adjustments based on fitness measurements derived from physical simulations. In Figure~\ref{fig:app_phys}, we demonstrate sewing pattern editing guided by body pressure analysis, including adjustments to the cuff (a), upper bodice (b), lower bodice (c), and collar (d). As shown in the examples, our system accurately identifies areas with excessive tension and adjusts the corresponding sewing patterns to enhance comfort while preserving the overall design.

\textbf{Generating New Garment Programs.} A major challenge in traditional parametric pattern-making is the need to abstract symbolic programs for new sewing patterns, which demands both advanced programming skills and pattern-making expertise. Design2GarmentCode addresses this by correlating GarmentCode grammar with LMMs' embedded pattern-making knowledge, enabling the automatic creation of new garment components. Figure~\ref{fig:app_new_code} shows a layered-skirt component generated by our DSL-GA, along with 3D garment representations demonstrating different design parameters, such as skirt length (c), number of layers (d), and layer differences such as length difference and ruffling factor (e). The results demonstrate that our system consistently produces high-quality garment components that meet professional standards, while significantly reducing the time and expertise required to create new sewing pattern programs.
\section{Conclusion}

Design2GarmentCode transforms multi-modal design concepts into precise sewing patterns using LMMs to synthesize parametric programs. It addresses challenges related to data requirements, computation, and the limited precision of neural network-based methods. 
The experimental results demonstrate the system's ability to capture design details while maintaining structural integrity and geometric precision in generated patterns. Despite these advantages, Design2GarmentCode currently cannot substantially alter GarmentCode’s underlying structure and logistics, which impacts generation quality due to inherent limitations in GarmentCode's design and modeling capabilities (Supp.~\ref{supp:limitations}).

\section*{Acknowledgments}
This work was supported by Key R\&D Program of Hangzhou (No.2024SDZ1A20). We sincerely thank Fuqi Wang for his valuable support in validation data collection, material modeling and animation.

{
    \small
    \bibliographystyle{ieeenat_fullname}
    \bibliography{main}

\begin{thebibliography}{67}
\providecommand{\natexlab}[1]{#1}
\providecommand{\url}[1]{\texttt{#1}}
\expandafter\ifx\csname urlstyle\endcsname\relax
  \providecommand{\doi}[1]{doi: #1}\else
  \providecommand{\doi}{doi: \begingroup \urlstyle{rm}\Url}\fi

\bibitem[Ass()]{Assyst}
Assyst smart.pattern macro.
\newblock \url{https://www.assyst.de/en/products/cad/index.html}.
\newblock Accessed: 2024-11-09.

\bibitem[Mod()]{Modaris}
Modaris: Smart and speedy patternmaking for fashion experts.
\newblock \url{https://www.lectra.com/en/fashion/products/modaris}.
\newblock Accessed: 2024-11-09.

\bibitem[Opt()]{Optitex}
Optitex: 2d/3d cad pattern design software.
\newblock \url{https://optitex.com/products/2d-and-3d-cad-software/}.
\newblock Accessed: 2024-11-09.

\bibitem[Val()]{Valentina}
Valentina: Open source pattern drafting software.
\newblock \url{https://valentinaproject.bitbucket.io/index.html}.
\newblock Accessed: 2024-11-09.

\bibitem[Sty(2024)]{Style3D}
{Style3D Studio}.
\newblock \url{https://www.linctex.com}, 2024.

\bibitem[Achiam et~al.(2023)Achiam, Adler, Agarwal, Ahmad, Akkaya, Aleman, Almeida, Altenschmidt, Altman, Anadkat, et~al.]{achiam2023gpt}
Josh Achiam, Steven Adler, Sandhini Agarwal, Lama Ahmad, Ilge Akkaya, Florencia~Leoni Aleman, Diogo Almeida, Janko Altenschmidt, Sam Altman, Shyamal Anadkat, et~al.
\newblock Gpt-4 technical report.
\newblock \emph{arXiv preprint arXiv:2303.08774}, 2023.

\bibitem[Bang et~al.(2021)Bang, Korosteleva, and Lee]{bang2021estimating}
Seungbae Bang, Maria Korosteleva, and Sung-Hee Lee.
\newblock Estimating garment patterns from static scan data.
\newblock \emph{Computer Graphics Forum}, 40\penalty0 (6):\penalty0 273--287, 2021.

\bibitem[Bao et~al.(2021)Bao, Miao, Gu, Liu, and Liu]{bao20213d}
Chen Bao, Yongwei Miao, Bingfei Gu, Kaixuan Liu, and Zhen Liu.
\newblock 3d interactive garment parametric pattern-making and linkage editing based on constrained contour lines.
\newblock \emph{International Journal of Clothing Science and Technology}, 33\penalty0 (5):\penalty0 696--723, 2021.

\bibitem[Bassamzadeh and Methani(2024)]{bassamzadeh2024comparative}
Nastaran Bassamzadeh and Chhaya Methani.
\newblock A comparative study of dsl code generation: Fine-tuning vs. optimized retrieval augmentation.
\newblock \emph{arXiv preprint arXiv:2407.02742}, 2024.

\bibitem[Carlier et~al.(2020)Carlier, Danelljan, Alahi, and Timofte]{carlier2020deepsvg}
Alexandre Carlier, Martin Danelljan, Alexandre Alahi, and Radu Timofte.
\newblock Deepsvg: A hierarchical generative network for vector graphics animation.
\newblock \emph{Advances in Neural Information Processing Systems}, 33:\penalty0 16351--16361, 2020.

\bibitem[Chen et~al.(2021)Chen, Tworek, Jun, Yuan, Pinto, Kaplan, Edwards, Burda, Joseph, Brockman, et~al.]{chen2021evaluating}
Mark Chen, Jerry Tworek, Heewoo Jun, Qiming Yuan, Henrique Ponde De~Oliveira Pinto, Jared Kaplan, Harri Edwards, Yuri Burda, Nicholas Joseph, Greg Brockman, et~al.
\newblock Evaluating large language models trained on code.
\newblock \emph{arXiv preprint arXiv:2107.03374}, 2021.

\bibitem[Chen et~al.(2022)Chen, Wang, Zhu, Liang, Torr, and Lin]{chen2022structure}
Xipeng Chen, Guangrun Wang, Dizhong Zhu, Xiaodan Liang, Philip Torr, and Liang Lin.
\newblock Structure-preserving 3d garment modeling with neural sewing machines.
\newblock \emph{Advances in Neural Information Processing Systems}, 35:\penalty0 15147--15159, 2022.

\bibitem[Chen et~al.(2023)Chen, Jiang, Liu, Huang, Fu, Chen, and Yu]{chen2023executing}
Xin Chen, Biao Jiang, Wen Liu, Zilong Huang, Bin Fu, Tao Chen, and Gang Yu.
\newblock Executing your commands via motion diffusion in latent space.
\newblock In \emph{Proceedings of the IEEE/CVF Conference on Computer Vision and Pattern Recognition}, pages 18000--18010, 2023.

\bibitem[Duan et~al.(2022)Duan, Liu, Wang, Zheng, Zhou, Chen, Hu, and Wang]{duan2022comprehensive}
Keyu Duan, Zirui Liu, Peihao Wang, Wenqing Zheng, Kaixiong Zhou, Tianlong Chen, Xia Hu, and Zhangyang Wang.
\newblock A comprehensive study on large-scale graph training: Benchmarking and rethinking.
\newblock \emph{Advances in Neural Information Processing Systems}, 35:\penalty0 5376--5389, 2022.

\bibitem[Dwivedi et~al.(2024)Dwivedi, Sun, Patel, Feng, and Black]{dwivedi_cvpr2024_tokenhmr}
Sai~Kumar Dwivedi, Yu Sun, Priyanka Patel, Yao Feng, and Michael~J. Black.
\newblock {TokenHMR}: Advancing human mesh recovery with a tokenized pose representation.
\newblock In \emph{IEEE/CVF Conference on Computer Vision and Pattern Recognition (CVPR)}, 2024.

\bibitem[Edward(1963)]{edward1963sketchpad}
Sutherland~Ivan Edward.
\newblock Sketchpad: A man-machine graphical communication system.
\newblock \emph{PhD Thesis, Massachusetts Institute of Technology}, 1963.

\bibitem[Ellis et~al.(2021)Ellis, Wong, Nye, Sabl{\'e}-Meyer, Morales, Hewitt, Cary, Solar-Lezama, and Tenenbaum]{ellis2021dreamcoder}
Kevin Ellis, Catherine Wong, Maxwell Nye, Mathias Sabl{\'e}-Meyer, Lucas Morales, Luke Hewitt, Luc Cary, Armando Solar-Lezama, and Joshua~B Tenenbaum.
\newblock Dreamcoder: Bootstrapping inductive program synthesis with wake-sleep library learning.
\newblock In \emph{Proceedings of the 42nd ACM Sigplan International Conference on Programming Language Design and Implementation}, pages 835--850, 2021.

\bibitem[Feng et~al.(2024)Feng, Lin, Dwivedi, Sun, Patel, and Black]{feng2024chatpose}
Yao Feng, Jing Lin, Sai~Kumar Dwivedi, Yu Sun, Priyanka Patel, and Michael~J Black.
\newblock Chatpose: Chatting about 3d human pose.
\newblock In \emph{Proceedings of the IEEE/CVF Conference on Computer Vision and Pattern Recognition}, pages 2093--2103, 2024.

\bibitem[Fried et~al.(2022)]{fried2022incoder}
Daniel Fried et~al.
\newblock Incoder: A generative model for code infilling and synthesis.
\newblock \emph{arXiv preprint arXiv:2204.05999}, 2022.

\bibitem[Gal et~al.(2024)Gal, Haviv, Alaluf, Bermano, Cohen-Or, and Chechik]{gal2024comfygen}
Rinon Gal, Adi Haviv, Yuval Alaluf, Amit~H Bermano, Daniel Cohen-Or, and Gal Chechik.
\newblock Comfygen: Prompt-adaptive workflows for text-to-image generation.
\newblock \emph{arXiv preprint arXiv:2410.01731}, 2024.

\bibitem[Guerrero et~al.(2022)Guerrero, Ha{\v{s}}an, Sunkavalli, M{\v{e}}ch, Boubekeur, and Mitra]{guerrero2022matformer}
Paul Guerrero, Milo{\v{s}} Ha{\v{s}}an, Kalyan Sunkavalli, Radom{\'\i}r M{\v{e}}ch, Tamy Boubekeur, and Niloy~J Mitra.
\newblock Matformer: A generative model for procedural materials.
\newblock \emph{arXiv preprint arXiv:2207.01044}, 2022.

\bibitem[He et~al.(2024)He, Yao, Zhang, Yu, Liu, and Xu]{he2024dresscode}
Kai He, Kaixin Yao, Qixuan Zhang, Jingyi Yu, Lingjie Liu, and Lan Xu.
\newblock Dresscode: Autoregressively sewing and generating garments from text guidance.
\newblock \emph{ACM Transactions on Graphics (TOG)}, 43\penalty0 (4):\penalty0 1--13, 2024.

\bibitem[Henighan et~al.(2020)Henighan, Kaplan, Katz, Chen, Hesse, Jackson, Jun, Brown, Dhariwal, Gray, et~al.]{henighan2020scaling}
Tom Henighan, Jared Kaplan, Mor Katz, Mark Chen, Christopher Hesse, Jacob Jackson, Heewoo Jun, Tom~B Brown, Prafulla Dhariwal, Scott Gray, et~al.
\newblock Scaling laws for autoregressive generative modeling.
\newblock \emph{arXiv preprint arXiv:2010.14701}, 2020.

\bibitem[Hu et~al.(2021)Hu, Shen, et~al.]{hu2021lora}
Edward~J. Hu, Yelong Shen, et~al.
\newblock Lora: Low-rank adaptation of large language models.
\newblock \emph{arXiv preprint arXiv:2106.09685}, 2021.

\bibitem[Hu et~al.(2022)Hu, He, Deschaintre, Dorsey, and Rushmeier]{hu2022inverse}
Yiwei Hu, Chengan He, Valentin Deschaintre, Julie Dorsey, and Holly Rushmeier.
\newblock An inverse procedural modeling pipeline for svbrdf maps.
\newblock \emph{ACM Transactions on Graphics (TOG)}, 41\penalty0 (2):\penalty0 1--17, 2022.

\bibitem[Jeong et~al.(2015)Jeong, Han, and Ko]{jeong2015garment}
Moon-Hwan Jeong, Dong-Hoon Han, and Hyeong-Seok Ko.
\newblock Garment capture from a photograph.
\newblock \emph{Computer Animation and Virtual Worlds}, 26\penalty0 (3-4):\penalty0 291--300, 2015.

\bibitem[Jones et~al.(2022)Jones, Walke, and Ritchie]{jones2022plad}
R~Kenny Jones, Homer Walke, and Daniel Ritchie.
\newblock Plad: Learning to infer shape programs with pseudo-labels and approximate distributions.
\newblock In \emph{Proceedings of the IEEE/CVF Conference on Computer Vision and Pattern Recognition}, pages 9871--9880, 2022.

\bibitem[Jones et~al.(2023)Jones, Guerrero, Mitra, and Ritchie]{jones2023shapecoder}
R~Kenny Jones, Paul Guerrero, Niloy~J Mitra, and Daniel Ritchie.
\newblock Shapecoder: Discovering abstractions for visual programs from unstructured primitives.
\newblock \emph{ACM Transactions on Graphics (TOG)}, 42\penalty0 (4):\penalty0 1--17, 2023.

\bibitem[Kang et~al.(2021)Kang, Oh, and Kim]{kang2021development}
Yeonghoon Kang, Jihyun Oh, and Sungmin Kim.
\newblock Development of parametric garment pattern design system.
\newblock \emph{International Journal of Clothing Science and Technology}, 33\penalty0 (5):\penalty0 724--739, 2021.

\bibitem[Korosteleva and Lee(2022)]{korosteleva2022neuraltailor}
Maria Korosteleva and Sung-Hee Lee.
\newblock Neuraltailor: Reconstructing sewing pattern structures from 3d point clouds of garments.
\newblock \emph{ACM Transactions on Graphics (TOG)}, 41\penalty0 (4):\penalty0 1--16, 2022.

\bibitem[Korosteleva and Sorkine-Hornung(2023)]{GarmentCode2023}
Maria Korosteleva and Olga Sorkine-Hornung.
\newblock {GarmentCode}: Programming parametric sewing patterns.
\newblock \emph{ACM Transaction on Graphics}, 42\penalty0 (6), 2023.

\bibitem[Korosteleva et~al.(2024)Korosteleva, Kesdogan, Kemper, Wenninger, Koller, Zhang, Botsch, and Sorkine-Hornung]{GarmentCodeData:2024}
Maria Korosteleva, Timur~Levent Kesdogan, Fabian Kemper, Stephan Wenninger, Jasmin Koller, Yuhan Zhang, Mario Botsch, and Olga Sorkine-Hornung.
\newblock {GarmentCodeData}: A dataset of 3{D} made-to-measure garments with sewing patterns.
\newblock In \emph{Computer Vision -- ECCV 2024}, 2024.

\bibitem[Li et~al.(2020)Li, Pan, Bousseau, and Mitra]{li2020sketch2cad}
Changjian Li, Hao Pan, Adrien Bousseau, and Niloy~J Mitra.
\newblock Sketch2cad: Sequential cad modeling by sketching in context.
\newblock \emph{ACM Transactions on Graphics (TOG)}, 39\penalty0 (6):\penalty0 1--14, 2020.

\bibitem[Li et~al.(2024{\natexlab{a}})Li, Li, Chen, Zhu, Guo, Lu, and Li]{li2024labyrinth}
Hong Li, Nanxi Li, Yuanjie Chen, Jianbin Zhu, Qinlu Guo, Cewu Lu, and Yong-Lu Li.
\newblock The labyrinth of links: Navigating the associative maze of multi-modal llms.
\newblock \emph{arXiv preprint arXiv:2410.01417}, 2024{\natexlab{a}}.

\bibitem[Li et~al.(2024{\natexlab{b}})Li, Guillard, and Fua]{li2024isp}
Ren Li, Beno{\^\i}t Guillard, and Pascal Fua.
\newblock Isp: Multi-layered garment draping with implicit sewing patterns.
\newblock \emph{Advances in Neural Information Processing Systems}, 36, 2024{\natexlab{b}}.

\bibitem[Li et~al.(2022)Li, Sekhon, Labash, et~al.]{li2022competition}
Yingjie Li, Arjun Sekhon, Brandon Labash, et~al.
\newblock Competition-level code generation with alphacode.
\newblock \emph{Science Advances}, 8\penalty0 (40):\penalty0 eabq4412, 2022.

\bibitem[Li et~al.(2024{\natexlab{c}})Li, Chen, Larionov, Sarafianos, Matusik, and Stuyck]{li2024diffavatar}
Yifei Li, Hsiao-yu Chen, Egor Larionov, Nikolaos Sarafianos, Wojciech Matusik, and Tuur Stuyck.
\newblock Diffavatar: Simulation-ready garment optimization with differentiable simulation.
\newblock In \emph{Proceedings of the IEEE/CVF Conference on Computer Vision and Pattern Recognition}, pages 4368--4378, 2024{\natexlab{c}}.

\bibitem[Li et~al.(2024{\natexlab{d}})Li, Wu, Liu, Wang, Dou, Ji, Zhang, Li, Lu, Tan, et~al.]{li2024isolated}
Yong-Lu Li, Xiaoqian Wu, Xinpeng Liu, Zehao Wang, Yiming Dou, Yikun Ji, Junyi Zhang, Yixing Li, Xudong Lu, Jingru Tan, et~al.
\newblock From isolated islands to pangea: Unifying semantic space for human action understanding.
\newblock In \emph{Proceedings of the IEEE/CVF Conference on Computer Vision and Pattern Recognition}, pages 16582--16592, 2024{\natexlab{d}}.

\bibitem[Liu et~al.(2024{\natexlab{a}})Liu, Xu, Yang, and Wang]{liu2024automatic}
Chen Liu, Weiwei Xu, Yin Yang, and Huamin Wang.
\newblock Automatic digital garment initialization from sewing patterns.
\newblock \emph{ACM Transactions on Graphics (TOG)}, 43\penalty0 (4):\penalty0 1--12, 2024{\natexlab{a}}.

\bibitem[Liu et~al.(2023{\natexlab{a}})Liu, Xu, Lin, Liang, and Yan]{liu2023sewformer}
Lijuan Liu, Xiangyu Xu, Zhijie Lin, Jiabin Liang, and Shuicheng Yan.
\newblock Towards garment sewing pattern reconstruction from a single image.
\newblock \emph{ACM Transactions on Graphics (SIGGRAPH Asia)}, 2023{\natexlab{a}}.

\bibitem[Liu et~al.(2024{\natexlab{b}})Liu, Li, Fang, Liu, You, and Lu]{liu2024primitive}
Siqi Liu, Yong-Lu Li, Zhou Fang, Xinpeng Liu, Yang You, and Cewu Lu.
\newblock Primitive-based 3d human-object interaction modelling and programming.
\newblock In \emph{Proceedings of the AAAI Conference on Artificial Intelligence}, pages 3711--3719, 2024{\natexlab{b}}.

\bibitem[Liu et~al.(2023{\natexlab{b}})Liu, Feng, Xiu, Liu, Paull, Black, and Schölkopf]{liu2023gshell}
Zhen Liu, Yao Feng, Yuliang Xiu, Weiyang Liu, Liam Paull, Michael~J. Black, and Bernhard Schölkopf.
\newblock Ghost on the shell: An expressive representation of general 3d shapes.
\newblock \emph{arXiv preprint arXiv:2310.15168}, 2023{\natexlab{b}}.

\bibitem[Nijkamp et~al.(2022)]{nijkamp2022codegen}
Erik Nijkamp et~al.
\newblock A conversational approach to code generation using pre-trained language models.
\newblock \emph{arXiv preprint arXiv:2203.13474}, 2022.

\bibitem[Para et~al.(2021{\natexlab{a}})Para, Bhat, Guerrero, Kelly, Mitra, Guibas, and Wonka]{para2021sketchgen}
Wamiq Para, Shariq Bhat, Paul Guerrero, Tom Kelly, Niloy Mitra, Leonidas~J Guibas, and Peter Wonka.
\newblock Sketchgen: Generating constrained cad sketches.
\newblock \emph{Advances in Neural Information Processing Systems}, 34:\penalty0 5077--5088, 2021{\natexlab{a}}.

\bibitem[Para et~al.(2021{\natexlab{b}})Para, Guerrero, Kelly, Guibas, and Wonka]{para2021generative}
Wamiq Para, Paul Guerrero, Tom Kelly, Leonidas~J Guibas, and Peter Wonka.
\newblock Generative layout modeling using constraint graphs.
\newblock In \emph{Proceedings of the IEEE/CVF international Conference on Computer Vision}, pages 6690--6700, 2021{\natexlab{b}}.

\bibitem[Pietroni et~al.(2022)Pietroni, Dumery, Falque, Liu, Vidal-Calleja, and Sorkine-Hornung]{pietroni2022computational}
Nico Pietroni, Corentin Dumery, Raphael Falque, Mark Liu, Teresa~A Vidal-Calleja, and Olga Sorkine-Hornung.
\newblock Computational pattern making from 3d garment models.
\newblock \emph{ACM Transactions on Graphics}, 41\penalty0 (4):\penalty0 157--1, 2022.

\bibitem[Ritchie et~al.(2023)Ritchie, Guerrero, Jones, Mitra, Schulz, Willis, and Wu]{ritchie2023neurosymbolic}
Daniel Ritchie, Paul Guerrero, R~Kenny Jones, Niloy~J Mitra, Adriana Schulz, Karl~DD Willis, and Jiajun Wu.
\newblock Neurosymbolic models for computer graphics.
\newblock \emph{Computer Graphics Forum}, 42\penalty0 (2):\penalty0 545--568, 2023.

\bibitem[Rong et~al.(2024)Rong, Grigorev, Wang, Black, Thomaszewski, Tsalicoglou, and Hilliges]{rong2024gaussian}
Boxiang Rong, Artur Grigorev, Wenbo Wang, Michael~J Black, Bernhard Thomaszewski, Christina Tsalicoglou, and Otmar Hilliges.
\newblock Gaussian garments: Reconstructing simulation-ready clothing with photorealistic appearance from multi-view video.
\newblock \emph{arXiv preprint arXiv:2409.08189}, 2024.

\bibitem[Seff et~al.(2021)Seff, Zhou, Richardson, and Adams]{seff2021vitruvion}
Ari Seff, Wenda Zhou, Nick Richardson, and Ryan~P Adams.
\newblock Vitruvion: A generative model of parametric cad sketches.
\newblock \emph{arXiv preprint arXiv:2109.14124}, 2021.

\bibitem[Shi et~al.(2020)Shi, Li, Ha{\v{s}}an, Sunkavalli, Boubekeur, Mech, and Matusik]{shi2020match}
Liang Shi, Beichen Li, Milo{\v{s}} Ha{\v{s}}an, Kalyan Sunkavalli, Tamy Boubekeur, Radomir Mech, and Wojciech Matusik.
\newblock Match: Differentiable material graphs for procedural material capture.
\newblock \emph{ACM Transactions on Graphics (TOG)}, 39\penalty0 (6):\penalty0 1--15, 2020.

\bibitem[Tian et~al.(2019)Tian, Luo, Sun, Ellis, Freeman, Tenenbaum, and Wu]{tian2019learning}
Yonglong Tian, Andrew Luo, Xingyuan Sun, Kevin Ellis, William~T Freeman, Joshua~B Tenenbaum, and Jiajun Wu.
\newblock Learning to infer and execute 3d shape programs.
\newblock \emph{arXiv preprint arXiv:1901.02875}, 2019.

\bibitem[Touvron et~al.(2023)Touvron, Lavril, Izacard, Martinet, Lachaux, Lacroix, Rozi{\`e}re, Goyal, Hambro, Azhar, et~al.]{touvron2023llama}
Hugo Touvron, Thibaut Lavril, Gautier Izacard, Xavier Martinet, Marie-Anne Lachaux, Timoth{\'e}e Lacroix, Baptiste Rozi{\`e}re, Naman Goyal, Eric Hambro, Faisal Azhar, et~al.
\newblock Llama: Open and efficient foundation language models.
\newblock \emph{arXiv preprint arXiv:2302.13971}, 2023.

\bibitem[Wang(2018)]{wang2018rule}
Huamin Wang.
\newblock Rule-free sewing pattern adjustment with precision and efficiency.
\newblock \emph{ACM Transactions on Graphics}, 37\penalty0 (4):\penalty0 1--13, 2018.

\bibitem[Wang et~al.(2019)Wang, Lin, Weissmann, Savva, Chang, and Ritchie]{wang2019planit}
Kai Wang, Yu-An Lin, Ben Weissmann, Manolis Savva, Angel~X Chang, and Daniel Ritchie.
\newblock Planit: Planning and instantiating indoor scenes with relation graph and spatial prior networks.
\newblock \emph{ACM Transactions on Graphics (TOG)}, 38\penalty0 (4):\penalty0 1--15, 2019.

\bibitem[Wang et~al.(2021)Wang, Wan, et~al.]{wang2021codet5}
Yue Wang, Fangxiang Wan, et~al.
\newblock Codet5: Identifier-aware unified pre-trained encoder-decoder models for code understanding and generation.
\newblock In \emph{Proceedings of the 2021 Conference on Empirical Methods in Natural Language Processing}, pages 8696--8708, 2021.

\bibitem[Wang et~al.(2018)Wang, Wu, Fratarcangeli, Tang, and Wang]{wang2018parallel}
Zhendong Wang, Longhua Wu, Marco Fratarcangeli, Min Tang, and Huamin Wang.
\newblock Parallel multigrid for nonlinear cloth simulation.
\newblock \emph{Computer Graphics Forum}, 37\penalty0 (7):\penalty0 131--141, 2018.

\bibitem[Wu et~al.(2021)Wu, Xiao, and Zheng]{wu2021deepcad}
Rundi Wu, Chang Xiao, and Changxi Zheng.
\newblock Deepcad: A deep generative network for computer-aided design models.
\newblock In \emph{Proceedings of the IEEE/CVF International Conference on Computer Vision}, pages 6772--6782, 2021.

\bibitem[Wu et~al.(2024)Wu, Li, Sun, and Lu]{wu2024symbol}
Xiaoqian Wu, Yong-Lu Li, Jianhua Sun, and Cewu Lu.
\newblock Symbol-llm: leverage language models for symbolic system in visual human activity reasoning.
\newblock \emph{Advances in Neural Information Processing Systems}, 36, 2024.

\bibitem[Xu et~al.(2022{\natexlab{a}})]{xu2022polycoder}
Kevin Xu et~al.
\newblock Polycoder: Multilingual code completion and generation.
\newblock \emph{arXiv preprint arXiv:2203.10290}, 2022{\natexlab{a}}.

\bibitem[Xu et~al.(2022{\natexlab{b}})Xu, Willis, Lambourne, Cheng, Jayaraman, and Furukawa]{xu2022skexgen}
Xiang Xu, Karl~DD Willis, Joseph~G Lambourne, Chin-Yi Cheng, Pradeep~Kumar Jayaraman, and Yasutaka Furukawa.
\newblock Skexgen: Autoregressive generation of cad construction sequences with disentangled codebooks.
\newblock \emph{arXiv preprint arXiv:2207.04632}, 2022{\natexlab{b}}.

\bibitem[Xu et~al.(2024{\natexlab{a}})Xu, Lambourne, Jayaraman, Wang, Willis, and Furukawa]{xu2024brepgen}
Xiang Xu, Joseph Lambourne, Pradeep Jayaraman, Zhengqing Wang, Karl Willis, and Yasutaka Furukawa.
\newblock Brepgen: A b-rep generative diffusion model with structured latent geometry.
\newblock \emph{ACM Transactions on Graphics (TOG)}, 43\penalty0 (4):\penalty0 1--14, 2024{\natexlab{a}}.

\bibitem[Xu et~al.(2024{\natexlab{b}})Xu, Zhang, Li, Han, and Lu]{xu2024humanvla}
Xinyu Xu, Yizheng Zhang, Yong-Lu Li, Lei Han, and Cewu Lu.
\newblock Humanvla: Towards vision-language directed object rearrangement by physical humanoid.
\newblock \emph{arXiv preprint arXiv:2406.19972}, 2024{\natexlab{b}}.

\bibitem[Xue et~al.(2024)Xue, Lu, Huang, Ouyang, and Bai]{xue2024genagent}
Xiangyuan Xue, Zeyu Lu, Di Huang, Wanli Ouyang, and Lei Bai.
\newblock Genagent: Build collaborative ai systems with automated workflow generation--case studies on comfyui.
\newblock \emph{arXiv preprint arXiv:2409.01392}, 2024.

\bibitem[Yang et~al.(2018)Yang, Pan, Amert, Wang, Yu, Berg, and Lin]{yang2018physics}
Shan Yang, Zherong Pan, Tanya Amert, Ke Wang, Licheng Yu, Tamara Berg, and Ming~C Lin.
\newblock Physics-inspired garment recovery from a single-view image.
\newblock \emph{ACM Transactions on Graphics (TOG)}, 37\penalty0 (5):\penalty0 1--14, 2018.

\bibitem[Yu et~al.(2023)Yu, Dou, Long, Lin, Li, Liu, Müller, Komura, Habermann, Theobalt, et~al.]{yu2023surf}
Zhengming Yu, Zhiyang Dou, Xiaoxiao Long, Cheng Lin, Zekun Li, Yuan Liu, Norman Müller, Taku Komura, Marc Habermann, Christian Theobalt, et~al.
\newblock Surf-d: High-quality surface generation for arbitrary topologies using diffusion models.
\newblock \emph{arXiv preprint arXiv:2311.17050}, 2023.

\bibitem[Zhang et~al.(2025)Zhang, Cai, Liu, Xu, Lu, and Li]{zhang2025take}
Mingyu Zhang, Jiting Cai, Mingyu Liu, Yue Xu, Cewu Lu, and Yong-Lu Li.
\newblock Take a step back: Rethinking the two stages in visual reasoning.
\newblock In \emph{European Conference on Computer Vision}, pages 124--141. Springer, 2025.

\bibitem[Zhou et~al.(2024)Zhou, Zhang, Ma, Shi, Liu, and Han]{zhou2024udiff}
Junsheng Zhou, Weiqi Zhang, Baorui Ma, Kanle Shi, Yu-Shen Liu, and Zhizhong Han.
\newblock Udiff: Generating conditional unsigned distance fields with optimal wavelet diffusion.
\newblock In \emph{Proceedings of the IEEE/CVF Conference on Computer Vision and Pattern Recognition}, pages 21496--21506, 2024.

\end{thebibliography}
}


\clearpage
\setcounter{page}{1}
\maketitlesupplementary

\section{Validations in Design2GarmentCode}
\label{supp:validation}
\subsection{Rule-based Validation}
\label{supp:validation:rule_based}
Rule-based validation primarily addresses issues of completeness and hallucination during the MMUA's generation process. With prompts generated by the DSL-GA containing over 100 questions, the MMUA often struggles to provide comprehensive answers in a single attempt. Additionally, due to the inherent hallucination tendencies of LLMs, some responses may fall outside the reasonable parameter range defined by GarmentCode. To mitigate these issues, we compare the MMUA's responses against a predefined complete question space to verify whether all questions have been adequately addressed before program synthesis. Each response is further validated to ensure it falls within GarmentCode's permissible parameter space. Questions with either missing or invalid answers are sent back to the MMUA for re-evaluation, with a maximum of two validation loops to refine the outputs.

\subsection{MMUA Design Comparison}
\label{supp:validation:comp_based}
\begin{figure}
    \centering
    \includegraphics[width=\linewidth]{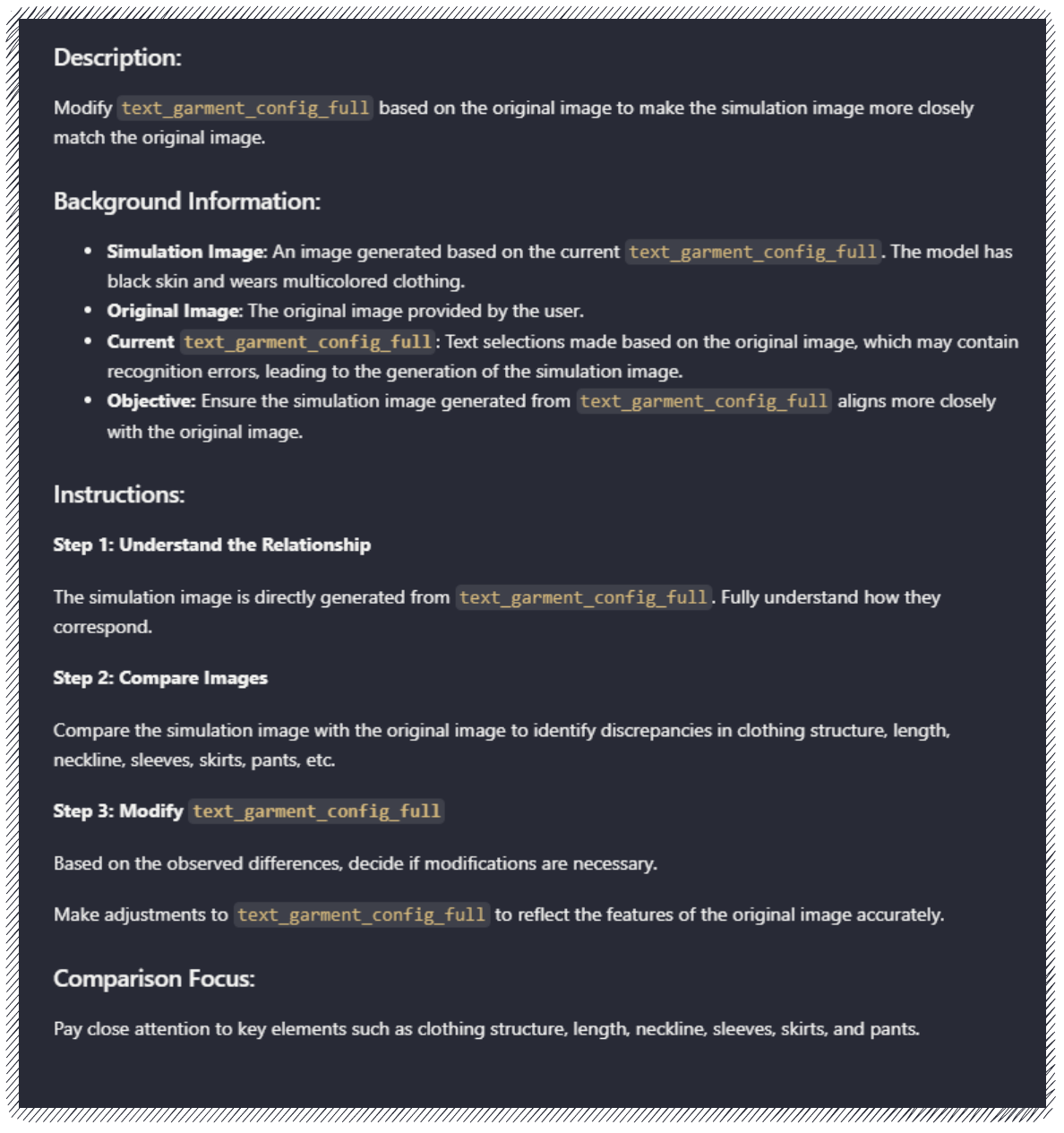}
    \caption{Prompt for MMUA during design comparison.}
    \label{fig:design_comp_prompt}
\end{figure}

During design comparison we ask the MMUA to compare the output design image versus the design input and propose modification suggestions to DSL-GA to edit the generated pattern. Design comparison is especially useful for image-guided generation, where the output design image is rendered from the draped garment mesh under a similar viewpoint to the input image, we use TokenHMR~\citep{dwivedi_cvpr2024_tokenhmr} to estimate the camera pose and rough human pose from the input design image. The prompt used for design comparison is given in Figure~\ref{fig:design_comp_prompt}.

\begin{figure*}
    \centering
    \includegraphics[width=\linewidth]{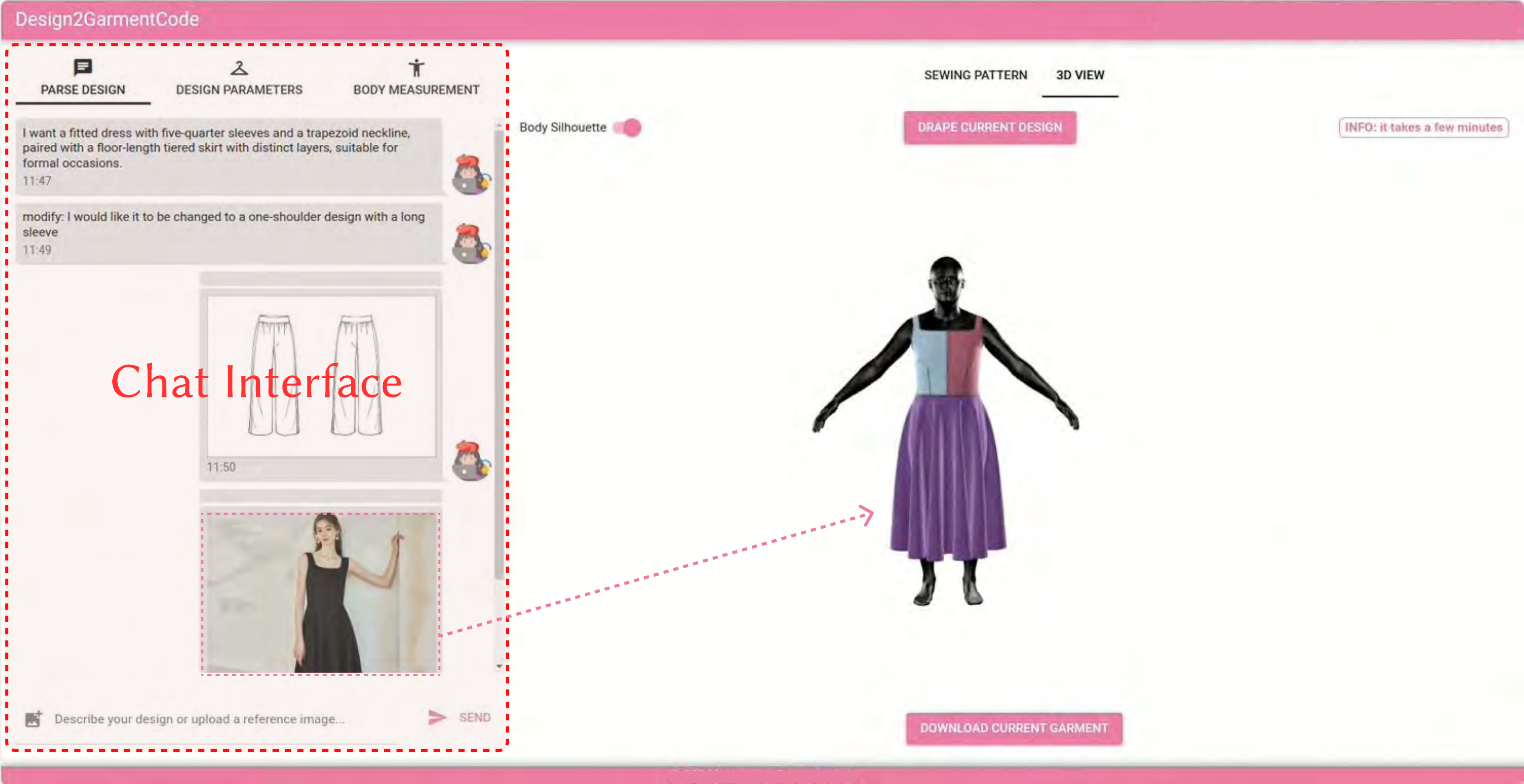}
    \caption{LMM-based interface for Design2GarmentCode, built upon the original GarmentCode GUI. The chat interface (left) allows users to provide natural language design descriptions or upload reference images or sketches, facilitating multi-modal design parsing into executable GarmentCode programs. We use the original GarmentCode execution engine to turn the generated program into 3D garments.}
    \label{fig:inference_gui}
\end{figure*}

\begin{figure*}
    \centering
    \includegraphics[width=\linewidth]{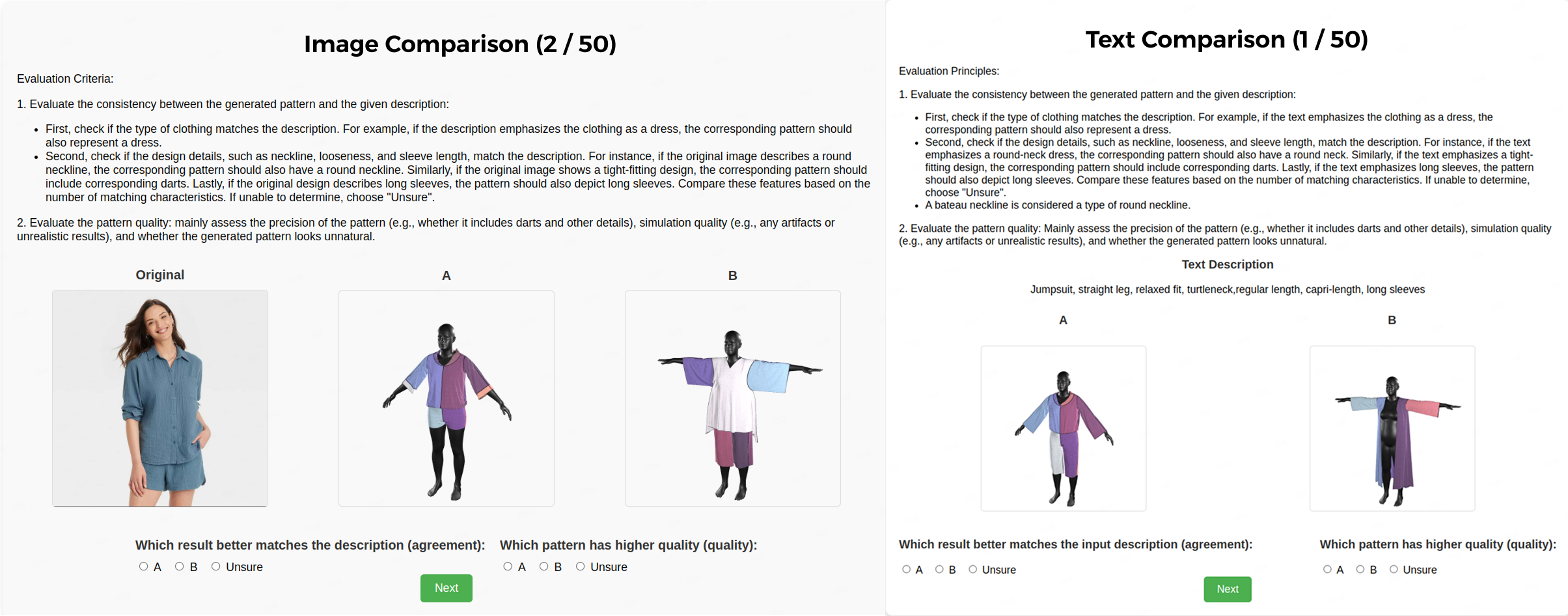}
    \caption{User study interface for evaluating sewing pattern generation quality. For each test input (Original image on the left for image-based evaluation or Text Description on the right for text-based evaluation), participants are presented with the simulation results of sewing patterns generated by Design2GarmentCode and a baseline method. Users are asked to evaluate the results based on two criteria: agreement with the input description and overall sewing pattern quality. If unsure, participants can select the "Unsure" option.}
    \label{fig:user_study_gui}
\end{figure*}

\section{Implementation Details}
\label{supp:implementation}
We use GPT-4V~\citep{achiam2023gpt} for \textbf{MMUA}, and an instruction tuned version of Llama-3.2-3B for \textbf{DSL-GA} ($\Gamma$). The following sections contains the detailed explanation for the finetuned DSL-GA (Supp.~\ref{supp:dsl_ga_ft}), and training details for the Projector $\Psi$ (Supp.~\ref{supp:train_proj}). 

\subsection{Finetuning DSL-GA $\Gamma_{ft}$}
\label{supp:dsl_ga_ft}
To optimize the trade-off between computational cost and generation quality, we use Llama-3.2-3B-Instruct\citep{touvron2023llama} as the base model for DSL-GA, fine-tuned over two epochs with LoRA (rank 16) and a learning rate of $5 \times 10^{-4}$ on a dataset with 583 hierarchically defined NL-DSL pairs from GarmentCode's public code repository. All code generation experiments were conducted on a single NVIDIA GTX 4090. For multi-modal understanding tasks, GPT-4V was employed as the designated agent.

\subsection{Training The Projector $\Psi$}
\label{supp:train_proj}
The projector $\Psi$ is trained on the GarmentCodeData~\citep{GarmentCodeData:2024} dataset, which comprises approximately 115,000 garment samples draped on a standard A-pose body. We generate initial design descriptions for each sample using GPT-4V or rule-based inverse mapping from the ground truth design parameters for the sample, for example
\begin{minted}{python}
    if design.shirt.length.v > 1.0: 
        return 'shirt__length__long'
\end{minted}
The token sequence length is fixed at $122$, which is equal to the number of design parameters in GarmentCode. The projection MLP and Transformer decoder are designed with feature dimensions of $128$. The MLP consists of $4$ intermediate layers, while the Transformer decoder includes $8$ layers. Training is conducted using the Adam optimizer with a learning rate of $5 \times 10^{-4}$, a batch size of $16$, and completed on a single NVIDIA GTX 4090 within $10$ hours.

Notably, although we adopt a decoder-only Transformer architecture similar to DressCode, our innovative approach of quantifying sewing patterns through design parameters proves to be significantly more efficient and scalable. Specifically, with DressCode’s quantization scheme, the token sequence length is calculated as:
\[
L_{seq}=N_p\times(N_e\times L_e + \|R\| + \|T\| + N_e\times \|S\|) + 2
\]
where $N_p$, $N_e$ denotes the maximum number of panels and edges respectively. $\|R\|=4$ is the length of rotation quaternions, and $\|T\|=3$ is the length of 3D translation vector. $\|S\|=4$ represents the per-edge stitching parameters containing a stitch tag and its existence indicator. $L_e$ represent the length of quantified edge vectors, which might be $6$ for cubic bezier curves and $4$ for quadratic bezier curves. Using GarmentCodeData as an example, to fully cover GarmentCode's modeling space, the required sequence length under DressCode’s method would be $13,951$, with $N_p=N_e=37,L_e=6$, which will cost $\approx 1.5h$ to generate a single sewing pattern using DressCode, while our token sequence length is fixed at $122$. 

\subsection{Inference Interface}
For more convenient inference, we build an intelligent chat-based interface integrated into the original GarmentCode~\citep{GarmentCode2023} GUI (Figure~\ref{fig:inference_gui}). The chat interface (left panel) enables users to provide natural language descriptions, upload reference images, or supply design sketches, facilitating multi-modal design parsing into GarmentCode-compliant programs which are then passed to the GarmentCode execution engine. The engine generates sewing patterns and 3D garment simulations (right panel). This interactive interface provides an intuitive environment for creating, editing, and refining sewing patterns, significantly improving accessibility for users without extensive expertise in parametric pattern-making. We provide a recording to demostrate the inference process in \href{demo.mp4}{demo.mp4}.


\begin{figure}
    \centering
    \includegraphics[width=\linewidth]{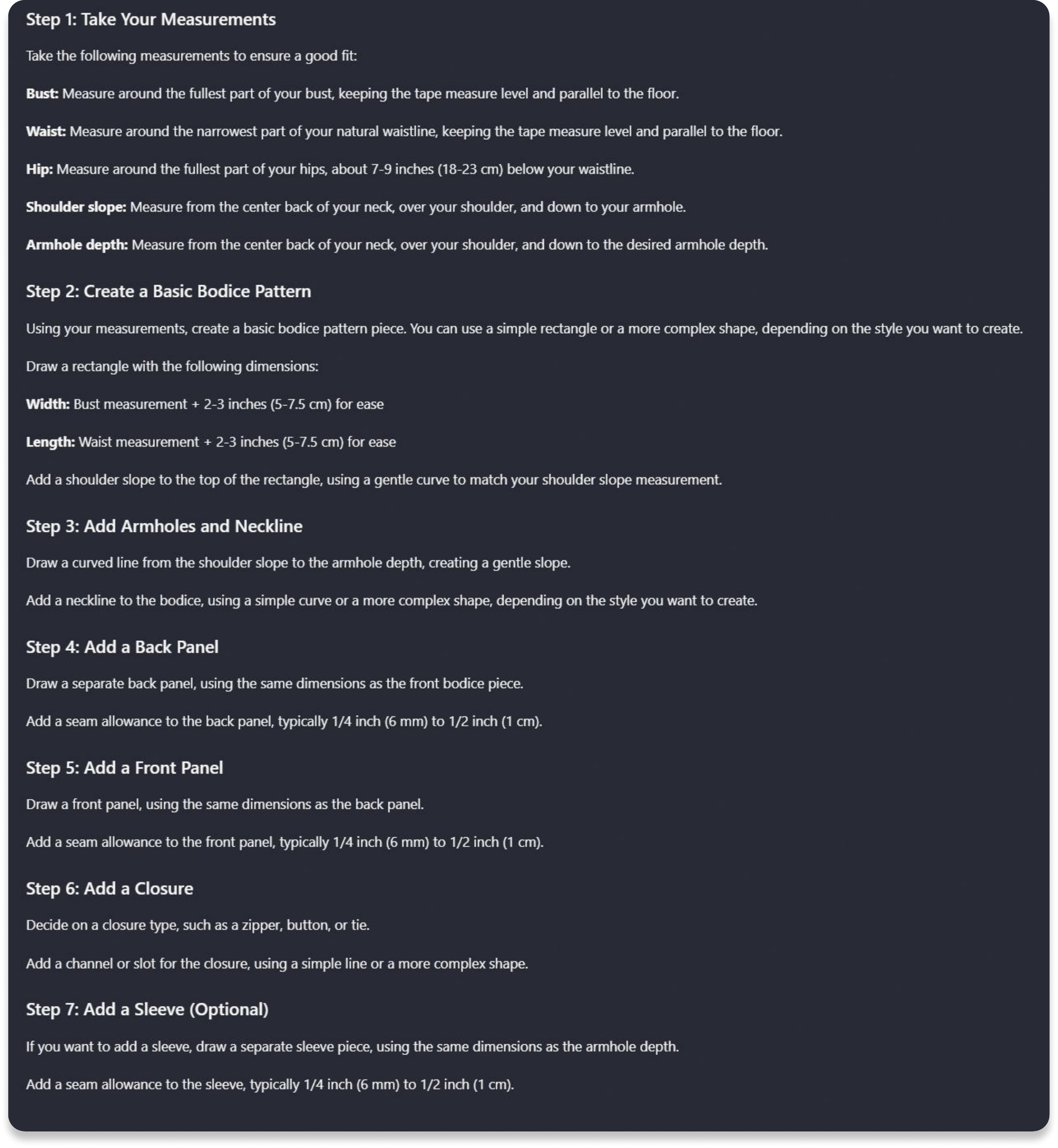}
    \caption{Example answers from Llama 3.2 3B when prompted with \textit{``How to draft a basic upper body bodice?''}.}
    \label{fig:pattern_drafting_test_llama}
\end{figure}

\begin{figure*}
    \centering
    \includegraphics[width=\linewidth]{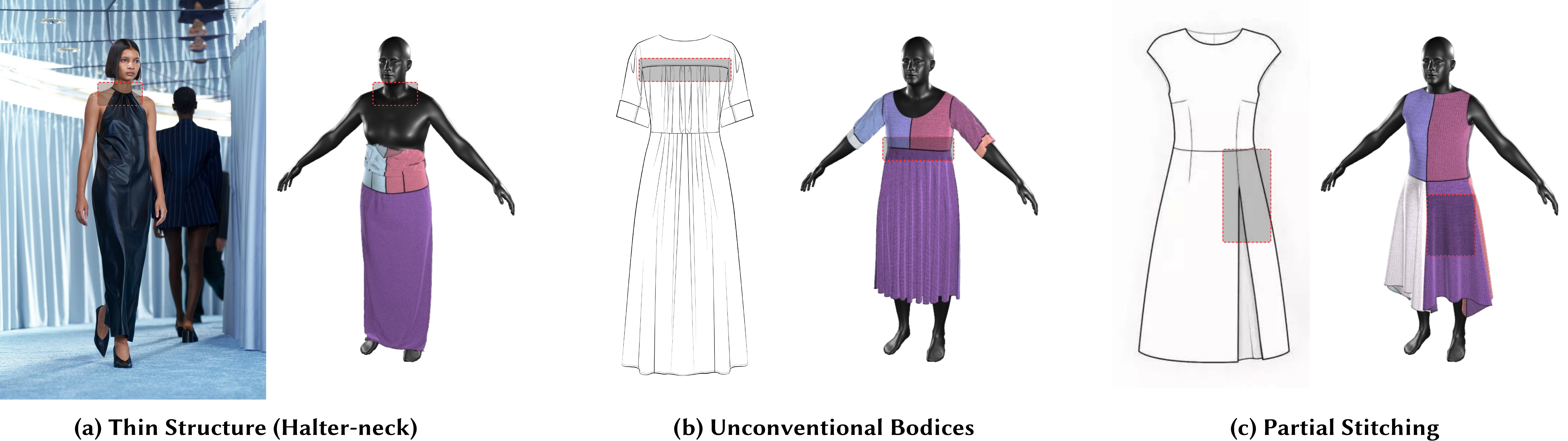}
    \caption{Limitations of Design2GarmentCode, including failed to modeling thin structures like halter-neck, unable to model unconventional bodices and stitching relationships are limited to one-to-one mapping.}
    \label{fig:limitations}
\end{figure*}

\subsection{User Study Interface}
To evaluate the quality of sewing pattern generation, we design a user study interface tailored for comparison (Figure~\ref{fig:user_study_gui}). For each test input—either an original image (for image-based evaluation) or a text description (for text-based evaluation)—the interface presents participants with simulated garment results generated by Design2GarmentCode and a baseline method. Participants assess the results based on two criteria: \textbf{Agreement}, which measures how well the generated patterns align with the input description, and \textbf{Aesthetic}, which evaluates the structural integrity and aesthetic appeal quality of the patterns. An "Unsure" option is available for cases where a clear preference cannot be determined, ensuring unbiased and flexible feedback.

\section{LMM Prompting Details}
\subsection{Pattern Drafting Test}
\label{supp:pattern_drafting}
As outlined in Sec.~\ref{sec:prog_learning}, a key prerequisite for Design2GarmentCode is the presence of embedded pattern-drafting knowledge in pre-trained large models. To assess this capability, we prompted models like O1-preview and LLama 3.2 3B Instruct with the question, "How to draft a basic upper body bodice?". These models produced step-by-step drafting instructions in natural language, including commands such as: "STEP 1: Take Your Measurements," and "STEP 2: Draw the Center Front Line, Draw the Shoulder Line, Draw the Armhole, Draw the Side Seam (Measure the distance from the underbust measurement and divide it by 4. Mark this distance from the armhole point down to the waist. Draw a vertical line to represent the side seam)." Figure~\ref{fig:pattern_drafting_test_llama} showcases sample outputs from Llama-3.2-3B Instruct which we used as baseline for DSL-GA.

\subsection{Prompting for MMUA}
\label{supp:prompts}
Based on different input design modalities and tasks, we assigned five specific tasks to the MMUA.

\textbf{Task 1:}
Identify the image, extract answers for each prompt question based on the image, and combine them to form the recognized clothing information. This task establishes the relationships between parameters and the questions corresponding to each parameter. It serves as the foundation for all subsequent tasks.

\textbf{Task 2:}
Generate clothing information based on text. Building on Task 1, this task generates clothing prototype information according to user preferences.

\textbf{Task 3:}
Retrieve existing clothing information and modify the clothing design according to the user's ideas.

\textbf{Task 4:}
Input stress test images along with the current clothing information from the text space. MMUA interprets the colors in the image as stress levels—red, yellow, or similar colors indicate areas that are too tight. MMUA dynamically adjusts the clothing information to reduce stress.

\textbf{Task 5:}
Compare previously generated clothing simulation images, their corresponding clothing information, and the original input image. Identify differences between the simulation and the original image, and dynamically adjust the clothing information to make the final simulation image more closely resemble the original.

\begin{figure}
    \centering
    \includegraphics[width=\linewidth]{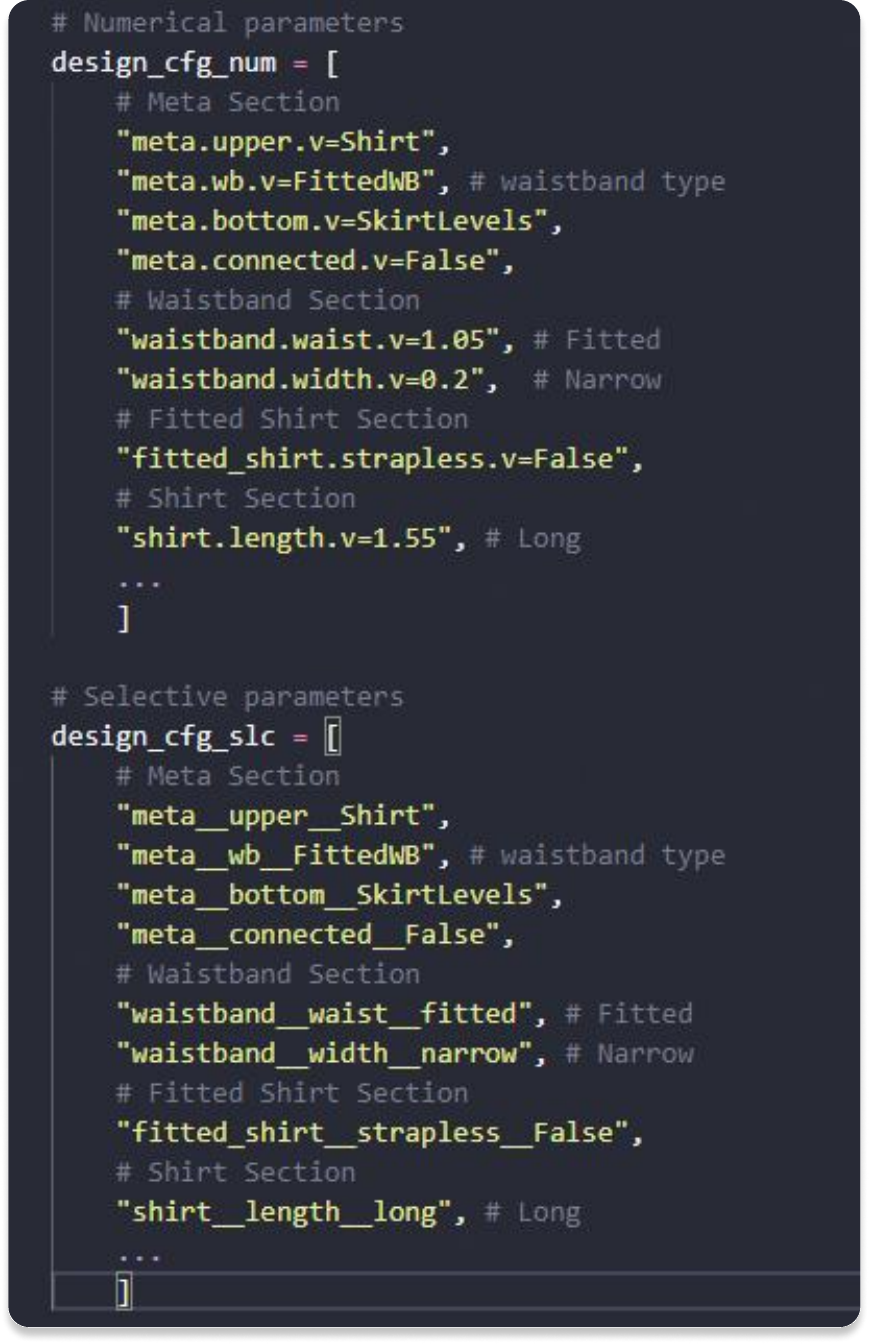}
    \caption{Example of original design configurations with numerical values and modified design configurations with only selective parameters.Example of original design configurations with numerical values and modified design configurations with only selective parameters.}
    \label{fig:param_comp}
\end{figure}

As discussed in Sec.~\ref{sec:prog_synthesis}, due to the probabilistic nature of LMMs, the MMUA struggles to accurately estimate numerical values in the design configuration. Therefore, we limit the MMUA's task to answering multiple-choice questions, with responses formatted as a selective parameter list. Fig.~\ref{fig:param_comp} illustrates example parameters before \textcolor[HTML]{3333FF}{(\texttt{design\_cfg\_num})} and after \textcolor[HTML]{3333FF}{(\texttt{design\_cfg\_slc})} modification. The complete prompt will be publicly available with Design2GarmentCode code base.

\section{Limitations}
\label{supp:limitations}
A limitation of Design2GarmentCode is its current inability to substantially modify GarmentCode's underlying structure and logic, which impacts the generation quality due to inherent constraints in GarmentCode's design and modeling capabilities. For example, the range of upper garment patterns is limited, making it difficult to model personalized segmentations (Figure~\ref{fig:limitations} (b)). Additionally, for designs like halter necks or strapless tops (Figure~\ref{fig:limitations} (a)), GarmentCode cannot model fine straps, leading to potential simulation failures. These constraints restrict the system's ability to accurately represent certain complex or customized garment designs.

\section{Additional Results}
In the following, we present additional experimental results in text-, image-, and sketch-guided pattern generation and highlight the versatility and effectiveness of our approach across various modalities. 

\begin{figure*}
    \centering
    \includegraphics[width=\linewidth]{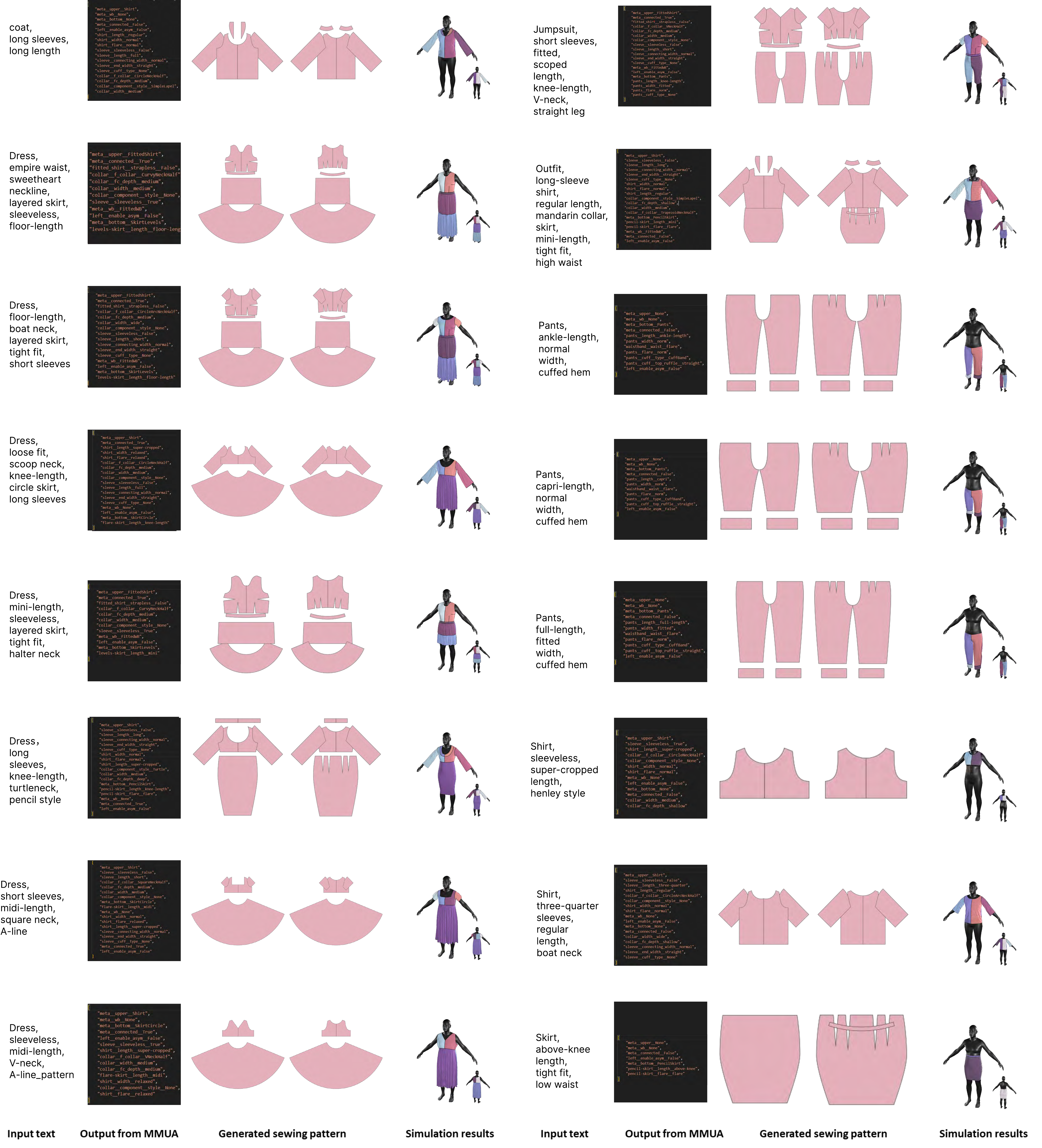}
    \caption{Additional Text-guided generation results. From left to right: input text; output from MMUA; generated sewing pattern; simulation results.}
    \label{fig:more_res_text}
\end{figure*}

\begin{figure*}
    \centering
    \includegraphics[width=\linewidth]{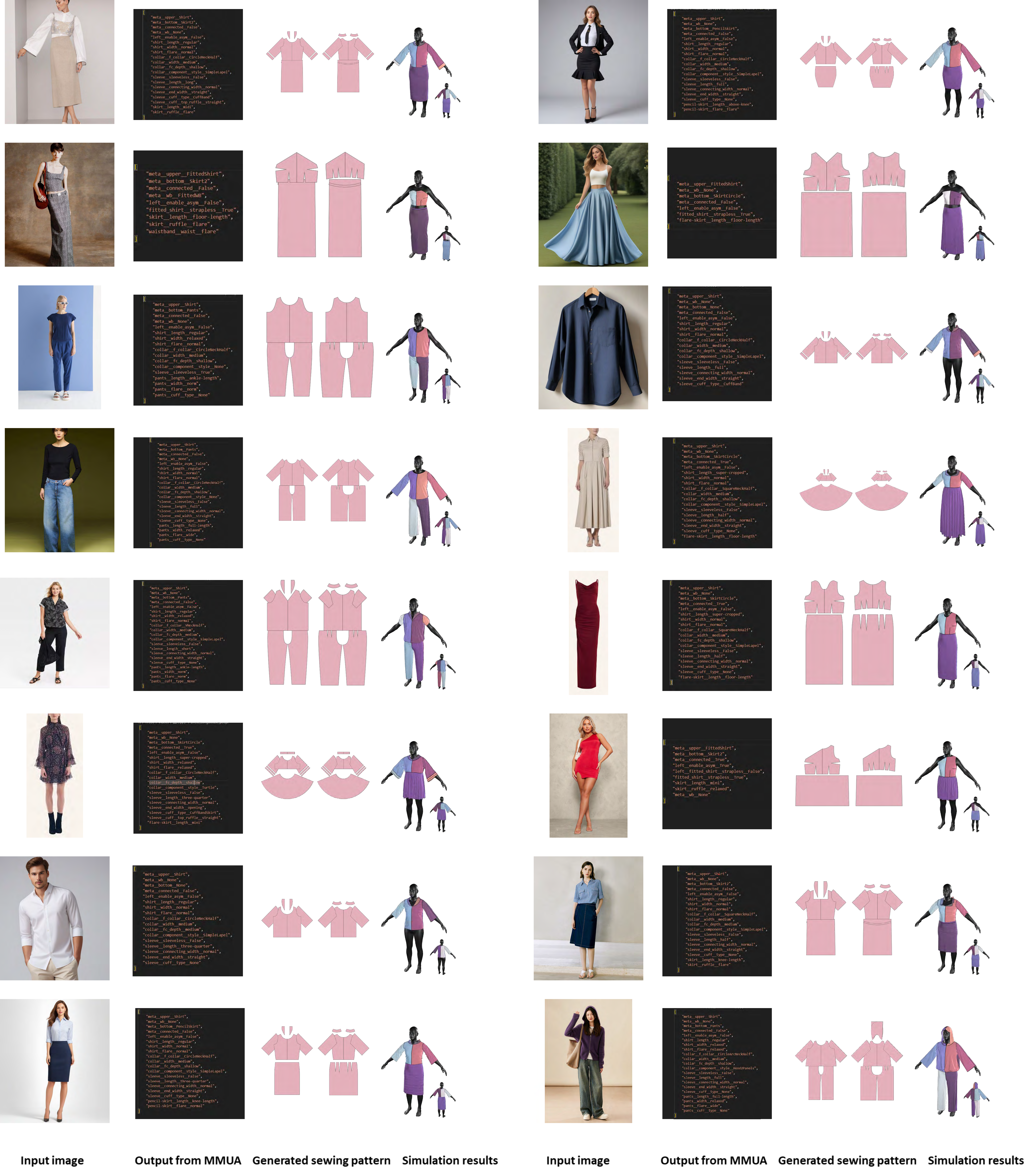}
    \caption{Additional Image-guided generation results. From left to right: input image; output from MMUA; generated sewing pattern; simulation results.}
    \label{fig:more_res_image}
\end{figure*}

\begin{figure*}
    \centering
    \includegraphics[width=\linewidth]{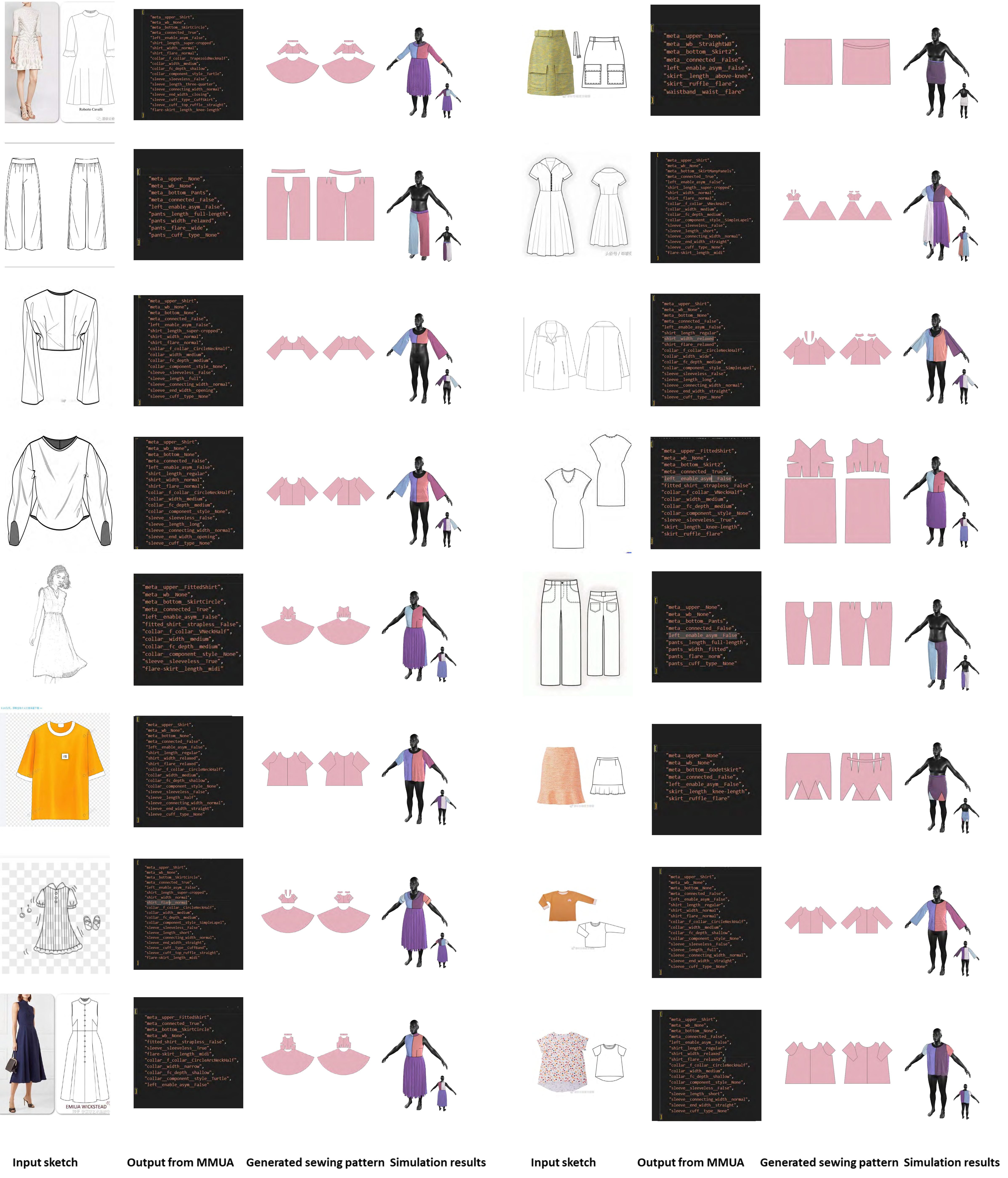}
    \caption{Additional Sketch-guided generation results. From left to right: input sketch; output from MMUA; generated sewing pattern; simulation results.}
    \label{fig:more_res_sketch}
\end{figure*}

\begin{figure*}
    \centering
    \includegraphics[width=\linewidth]{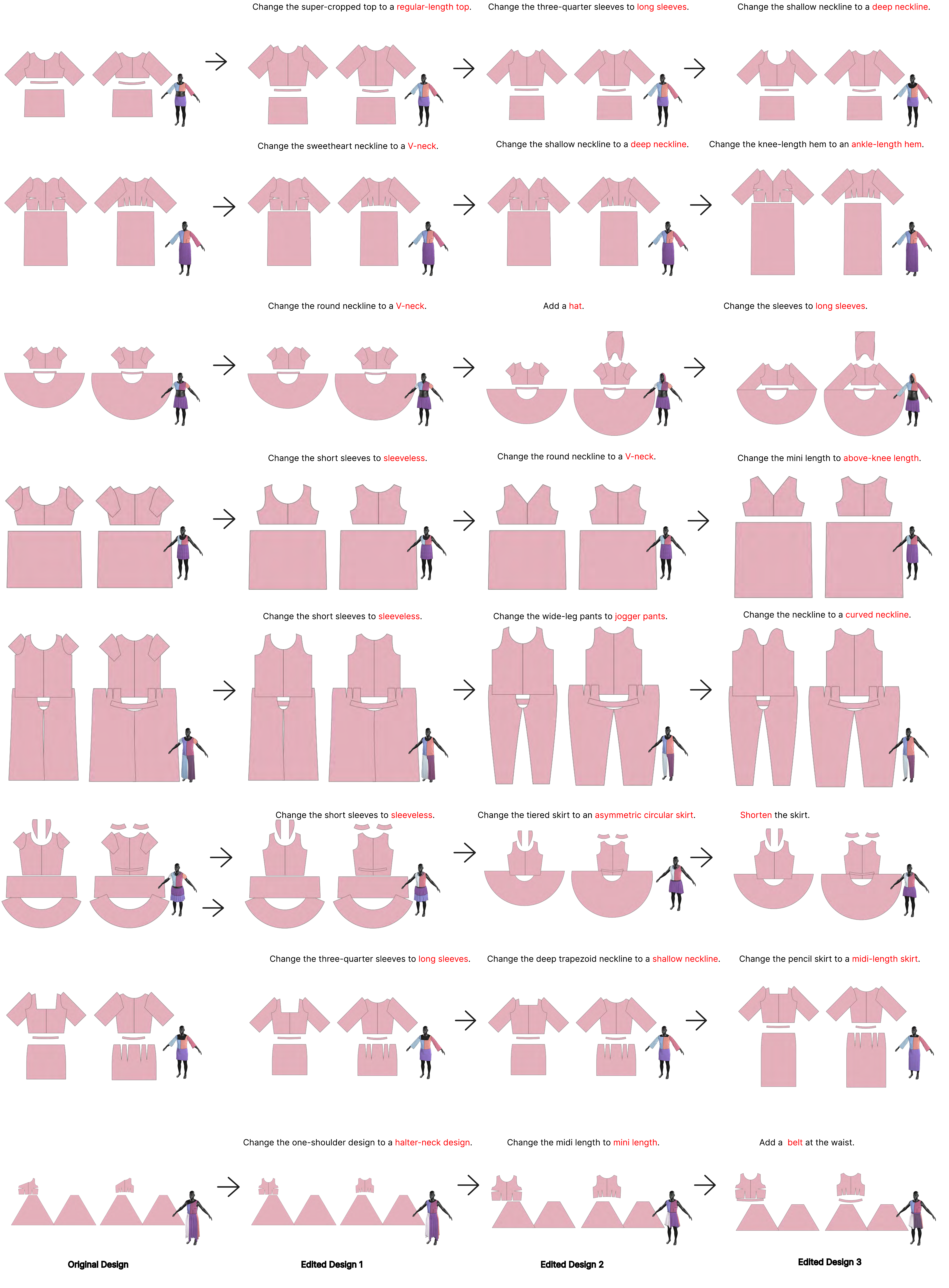}
    \caption{Additional sewing pattern editing results.Starting from the original sewing pattern on the far left, the system applies user instructions to edit the pattern. The left side of each arrow represents the original pattern, while the right side displays the edited result.}
    \label{fig:more_res_editing}
\end{figure*}

\end{document}